\documentclass[twocolumn,showkeys,        
               preprintnumbers,     
               aps,                 
               prd,          	    
               letterpaper,             
               superscriptaddress,      
               nofootinbib,         
               tightenlines,        
               floats,floatfix      
               ]{revtex4}
\usepackage{dcolumn}   
\usepackage{bm}        
\usepackage{float}
\usepackage{ulem}
\usepackage{lipsum}
\usepackage{graphicx,color}
\usepackage{latexsym}
\usepackage{amsmath,amssymb}        
\usepackage[colorlinks=true,linkcolor=blue,citecolor=blue]{hyperref}
\usepackage{mathrsfs}
\usepackage{comment}
\usepackage{soul}
\definecolor{purple}{rgb}{0.58,0.0,0.83}
\definecolor{orange}{rgb}{1,0.5,0}
\DeclareSymbolFontAlphabet{\mathrsfs}{rsfs}
\DeclareMathAlphabet{\mathcal}{OMS}{cmsy}{m}{n}

\begin{document}


\title{Solving the Schr\"odinger Poisson System using the coordinate Adaptive Moving Mesh method}


\author{Erick Munive-Villa}
\affiliation{Facultad de Ciencias F\'isico-Matem\'aticas, Ciudad Universitaria, Benem\'erita Universidad Aut\'onoma de Puebla, Av. San Claudio SN, Col. San Manuel, Puebla, M\'exico. }
\affiliation{Centro Internacional de F\'isica Fundamental, BUAP
\protect\\Ciudad Universitaria, Puebla, Puebla, M\'exico }

\author{Jessica N. L\'opez-S\'anchez}
\affiliation{Facultad de Ciencias F\'isico-Matem\'aticas, Ciudad Universitaria, Benem\'erita Universidad Aut\'onoma de Puebla, Av. San Claudio SN, Col. San Manuel, Puebla, M\'exico. }
\affiliation{Centro Internacional de F\'isica Fundamental, BUAP
\protect\\Ciudad Universitaria, Puebla, Puebla, M\'exico }

\author{Ana A. Avilez-López}
\email{ana.avilezlopez@correo.buap.mx}
\affiliation{Facultad de Ciencias F\'isico-Matem\'aticas, Ciudad Universitaria, Benem\'erita Universidad Aut\'onoma de Puebla, Av. San Claudio SN, Col. San Manuel, Puebla, M\'exico. } 
\affiliation{Centro Internacional de F\'isica Fundamental, BUAP
\protect\\Ciudad Universitaria, Puebla, Puebla, M\'exico }

\author{F. S. Guzm\'an}
\affiliation{
	      Instituto de F\'{\i}sica y Matem\'{a}ticas, Universidad
              Michoacana de San Nicol\'as de Hidalgo. Edificio C-3, Cd.
              Universitaria, 58040 Morelia, Michoac\'{a}n,
              M\'{e}xico.}


\date{\today}


\begin{abstract}
In this paper, we implement the Adaptive Moving Mesh method (AMM) to the solution of initial value problems involving the Schr\"odinger equation, and more specifically the Schr\"odinger-Poisson system of equations. This method is based on the solution of the problem on a discrete domain, whose resolution is coordinate and time-dependent, and allows to dynamically assign numerical resolution in terms of desired refinement criteria. We apply the method to solve various test problems involving stationary solutions of the SP system, and toy scenarios related to the disruption of subhalo s made of ultralight bosonic dark matter traveling on top of host galaxies.
\end{abstract}


\keywords{Schr\"odinger equation -- numerical methods-- mesh adaptation -- bosonic dark matter.}


\maketitle

\section{Introduction}
\label{sec:intro}
The possibility of scalar fields as dark matter has gained great interest in the community during the last decade due to its interesting phenomenological implications. Particularly, scalar field theories describing ultralight axion-like particles with masses around $10^{-22}eV$ have turned appealing owing to their lack of small scale problems, such as the missing satellites problem and the halo core-cusp problem, because of the large de Broglie length of such light particle of the order of kiloparsecs (e.g. \cite{Matos:1999,Matos-Urena:2000,Hu:2000,Hui:2016,Sahni:2000,Arbey2001a}). Particularly interesting is that the accumulation of bosons assembles macroscopic coherent states corresponding to a Bose-Einstein condensate that may play the role of dark matter halos as described in recent reviews (e.g. \cite{Hui:2021tkt,Lee:2017, Hui:2016,Suarez:2013}).

The analysis of the model includes the study of evolution and formation of structures, which needs the use of large-scale numerical simulations. The regime where this analysis happens is that in which the dynamic of the bosonic gas is ruled by the Gross-Pitaevskii-Poisson (GPP) system, where the parameter order obeys the Gross-Pitaevskii equation for the Bose gas subject to the gravitational potential sourced by the boson cloud itself. This analysis includes studies of structure formation and formation of universal density profiles (e. g. \cite{Schive:2017biq,Schive:2014dra, Schive:2014hza,Mocz:2017wlg,Mocz:2018,mocz19,mocz19b, Schwabe:2016}). 
{Given that processes of structure formation involve highly non-linear physics, these studies required developing codes to carry out simulations}, usually codes initially designed to study the structure formation within the CDM paradigm were adapted to include the Bosonic dark matter model in the fuzzy regime, for example, ENZO \cite{Enzo}, RAMSES \cite{Ramses}, AxioNyx \cite{Schwabe:2020_AxioNyx} and GADGET \cite{axionGad}. 

Interaction between a few structures need also the numerical solution of the GPP system that helps to study the universal properties of binary mergers and relaxation processes (e. g. \cite{BernalGuzman2006,Schwabe:2016,Guzman:2018evm,GuzmanAvilez2019,GuzmanGlezRios2021, Veltmaat:2018dfz}). Even the analysis of single structure scales involves numerical simulations describing the relaxation processes, for example the gravitational cooling \cite{GuzmanUrena2006,Seidel-Suen:1990,Shapiro2021}, or possible galactic halos with a specific structure, like a vortical solution to the GPP system, their stability and impact on galactic scale dynamics \cite{Shapiro2012,Guzman2014,LamHui2021,Shapiro2021b,Nicola2021}, or 
deformation of the core solitons making up the bosonic haloes due to tidal effects or rotation due to interactions in many-body systems \cite{Du:2018qor}. 

Explicitly, the Gross-Pitaevskii-Poisson system of equations that rules the dynamics in all these scenarios reads

\begin{eqnarray}\label{eq:schro}
i\hbar \frac{\partial \Psi}{\partial t}&=&-\frac{\hbar^2}{2m}\nabla^2 \Psi+mV \Psi, \\\label{eq:poisson}
\nabla^2 V &=& 4\pi G|\Psi|^2, 
\end{eqnarray}

\noindent where $m$ is the mass of a boson, $\Psi$ is an order parameter in the mean-field approximation at zero temperature of the Bose gas, subject to the trap of the gravitational potential $V$ sourced by the gas ground-state occupation-number density $|\psi|^2$ itself \cite{GuzmanUrena2004}. This system defines an Initial Value Problem (IVP) that is solved using a garden variety of numerical methods, for example, directly \cite{BernalGuzman:2006,UrenaBernal:2010,Schive:2014dra} or a Madelung transformed version of these equations (e.g. \cite{Li:2018kyk,Schwabe:2016, Matos:2019}), which is a hydrodynamical version of the GPP system of equations. Numerical methods vary from one code to another, some of them using finite differences, some others finite volume methods for the hydrodynamical version \cite{Mocz:2018}, some others use spectral methods \cite{Edwards:2018,Mocz:2017wlg} and finally Lagrangian methods inherited from Smoothed Particle Hydrodynamics \cite{axionGad,Enzo}.

Depending on the degree of detail and precision required in each case, three-dimensional codes that solve these equations use refinement strategies in order to optimize the computational resources with the aim of retaining precision in the regions where it is needed, for example in structure formation simulations \cite{Schive:2014dra,Mocz:2017wlg}, or in the simulations of local interaction of fluctuations (e.g. \cite{Schwabe:2016,Schwabe:2020_AxioNyx}) different spatial scales need different numerical resolution.

This is the reason why in this paper we add another possibility to this end. We apply the Adaptive Mesh Moving method (AMM) based on coordinate transformations, to the solution of problems involving the Schr\"odinger equation and the GPP system. These methods have been previously proposed for solving a whole class of partial differential equations, including astrophysics, for example in the early years of the binary black hole simulations prior to the use of AMR \cite{FishEye2000}.

The physical motivation of this paper is to describe in detail a method that can be helpful in the simulation of fuzzy dark matter dynamics at the scale of galactic halos. This is why in this paper we describe the implementation of the AMM method to solve the IVP associated with Schr\"odinger equation and especially the GPP system of equations above.

The paper is organized as follows. In Section \ref{sec:AMM} we describe the AMM method and in Section \ref{sec:GPP} the application to the GPP system. In Section \ref{sec:specificproblems} we present the strategy to implement the method to problems involving Schr\"odinger equation. Finally, in Section \ref{sec:comments} we draw some final comments.

\section{Adaptive Moving Mesh}
\label{sec:AMM}

For a general description of the method, consider an Initial Value Problem (IVP) formulated within the domain $\Omega \subset \mathbb{R}^3 \times t\in [0,t_f]\subset \mathbb{R}$, for the unknown $u=u({\bf x},t),~{\bf x}\in \mathbb{R}^3$ whose evolution equation reads:

\begin{equation}
    u_t +\nabla\cdot \mathbf{f} = \nabla\cdot(a\nabla u)+s,\label{eq:generalIVP}
\end{equation}

\noindent with appropriate initial conditions $u({\bf x},0)$ and boundary conditions $u(\partial \Omega,t)$ for $u$, and known functions $a$, $s$ and ${\bf f}$. The solution can be integrated using a discrete version of the problem defined on a discrete domain and then using an evolution method, assuming specific finite differentiation schemes in order to approximate spatial operators acting on fluxes, the parabolic term and the sources. 

The use of these methods needs the definition of a discrete domain. Assuming that $\Omega$ is a box described in Cartesian coordinates $\Omega=[x_{min},x_{max}]\times[y_{min},y_{max}]\times[z_{min},z_{max}]$, a simple discrete domain is the set of points $\Omega_P=\{(x_i,y_j,z_k)\in \Omega ~|~ x_i=x_{min}+i\Delta x, y_j=y_{min}+j\Delta y,z_k=z_{min}+k\Delta z\}$, where $i=0,...,N_x$, $j=0,...,N_y$ and $k=0,...,N_z$ are integer labels and $\Delta x=(x_{max}-x_{min})/N_x$, $\Delta y = (y_{max}-y_{min})/N_y$, $\Delta z=(z_{max}-z_{min})/N_z$ are the spatial resolutions of $\Omega_d$. The discretization of time is commonly defined as a function of  the spatial resolution, as the set of values $t^n\in[0,t_f]$ such that $t^n=n\Delta t$, where $\Delta t=CFL\min(\Delta x^{\alpha},\Delta y^{\alpha}, \Delta z^{\alpha})$, where $\alpha=1,2$ for only hyperbolic equations $a=0$ or parabolic $a\ne 0$ respectively and $CFL$ is the Courant-Friedrichs-Levy factor. In the case of Schr\"odinger equation $\alpha=2$.

Mesh Refinement is motivated by the need for accuracy in certain regions of the domain where the solution function $u$ is possibly developing structure or important features to the problem in turn. This method is implemented by defining a subset of interest $\Omega_s\subset \Omega$, possibly a box as well, with higher resolution $\Delta x_s,~\Delta y_s,~\Delta z_s$ where more accuracy is used to solve the problem in this local domain. Appropriate boundary conditions for $u$ at the inter-resolution boundary $\partial \Omega_s$ are implemented in terms of the values of $u$ in $\Omega_d$ using the well-known box in box method designed by Berger and Oliger \cite{berger1984amr}. As long as the refined domains remain fixed in space the method is usually called Fixed Mesh Refinement (FMR) and we will use it here for comparison with the AMM. Other refinement strategies allow these subdomains to move in time, which leads to the moving boxes method being well used in general relativistic simulations (e. g. \cite{MovingBoxes2008}). 

Although the AMM method is also motivated by the need for accuracy in certain regions of the spatial domain $\Omega_d$ it uses a different approach. Instead of defining new discrete sub-domains with higher resolutions as Mesh Refinement methods do, it redefines the equation associated with the IVP in terms of new coordinates. 

The new coordinates identify points of $\Omega_P$ with the normalized box called logical domain $\Omega_L=[0,1]^3$, whose discrete version is the uniformly distributed set of points $\Omega_{L}=\{(\xi_i,\eta_j,\kappa_k) ~|~ \xi_i=i\Delta \xi,\eta_j=j\Delta\eta,\kappa_k=k\Delta\kappa \}$, where $i=0,...,N_{\xi}$, $k=0,...,N_{\kappa}$ and $\Delta\xi=1/N_{\xi}$, $\Delta\eta=1/N_{\eta}$, $\Delta_{\kappa}=1/N_{\kappa}$ are the spatial resolutions of this Logical Domain. For our purposes we use $\Delta \xi=\Delta \eta=\Delta \kappa$.

On the other hand, we keep in mind that the physical discrete domain where the original problem has been defined for Eq. (\ref{eq:generalIVP}) is $\Omega_{P}$ ($P$ stands for physical) will no longer be uniform. Assume that more resolution is needed in the subdomain $U\subset \Omega$ due to a given refinement criterion. The AMM method would assign a high resolution to the physical discrete domain in the region of $U$ using a coordinate transformation. 
The discrete version of Eq. (\ref{eq:generalIVP}) is solved in the logical domain $\Omega_{L}$ where the spatial resolution is uniform and the solution transformed back to the physical domain $\Omega_{P}$ as we describe in detail below. The transformation between logical and physical domains can be dynamical.

\subsection{Physical to Logical Domain Transformation}\label{subsect:AMMPDE}

The general coordinate transformation between the two domains is a differentiable and invertible application

\begin{equation}
    \mathcal{T}: \Omega_{L}\subset \mathbb{R}^3 \rightarrow \Omega_{P}\subset \mathbb{R}^3 ,\nonumber
\end{equation}

\noindent that takes ${\bf \xi} \in \Omega_{L}$ and delivers ${\bf x}={\bf x}({\bf \xi})$, where $\mathbf{\xi}\in \Omega_{L}$ and $\mathbf{x}\in\Omega_{P}$. The inverse of $\mathcal{T}$ reads 

\begin{equation}
\mathcal{T}^{-1}: \Omega_{P} \rightarrow \Omega_{L},\nonumber
\end{equation}

\noindent which delivers $\mathbf{\xi}=\mathbf{\xi}(\mathbf{x})\in \Omega_L$. By means of $\mathcal{T}$, the unknown function $u$ can be written in terms of the logical space coordinates $\hat{u}$ as $\hat{u}(\mathbf{\xi},t)=u(\mathbf{\xi}(\mathbf{x},t))$.

The transformation and its inverse are constructed based on a variational method that extremises the functionals $I[\mathbf{\xi}]$ and $\hat{I}[\mathbf{x}]$ given by 

\begin{eqnarray}\label{eq:func1}
I[\mathbf{\xi}] &=& \int_{\Omega_{P}} F(\mathcal{J}^{-1},\mathbf{\xi},\mathbf{x})d\mathbf{x},\\\label{eq:func2}
\hat{I}[\mathbf{x}] &=& \int_{\Omega_{L}} \hat{F}(\mathcal{J},\mathbf{\xi},\mathbf{x})d\mathbf{\xi},
\end{eqnarray}

\noindent where $\mathcal{J}=\mathbf{\nabla}_{\xi}\mathbf{x}$ and $\mathcal{J}^{-1}=\mathbf{\nabla}_{\bf x}\mathbf{\xi}$ are the Jacobians associated to $\mathcal{T}$ and $\mathcal{T}^{-1}$ respectively, and $\nabla$ and $\nabla_{\xi}$ are the gradient operators with respect to the physical and logical coordinates. A common choice for $F$ has the form

\begin{equation}
F(\mathcal{J}^{-1},\xi,\mathbf{x})=F_1(\rho,\beta)+F_2(\rho,J),
\end{equation}

\noindent where $\rho(\mathbf{x})=\sqrt{\det(M(\mathbf{x}))}$ is the {\it mesh density function}, whereas $M=M(\mathbf{x})$ is the {\it monitor function}, and $\beta$ is written in terms of $\mathcal{J}^{-1}$ and $M$ reads

\begin{equation}
\beta=\sum_i(\nabla\xi_i)^TM^{-1}\nabla\xi_i.
\end{equation}

\noindent For the present work, we consider the particular case where

\begin{equation}
M = \mathbb{I}_n \omega,\nonumber
\end{equation}

\noindent so that $\rho = |\omega|^{3/2}$ and

\begin{equation}
    \hat{F}(\mathcal{J},\xi,\mathbf{x})=\rho\sum_i (\nabla_{\xi} x_i)^\text{T}\cdot\nabla_\xi x_i,
\end{equation}

\noindent where $\rho = |\omega|^{3/2}$ is considered to have a particular dependence on $\mathbf{x}$ via an arbitrary differentiable function $v$ that defines the concentration of resolution as a function of physical coordinates, so that $\omega=\omega(\mathbf{x},v)$. The Euler-Lagrange equations for the Logical domain resulting from the minimization of (\ref{eq:func2}) reduce to

\begin{equation}\label{eq:ELEQ}
\nabla_\xi \cdot(\rho \nabla_\xi x_i) = 0,
\end{equation}

\noindent where $i=$ 1, 2, 3 labels the coordinates $x$, $y$, $z$ respectively. This is a set of differential equations that need to be solved in order to obtain the transformation between the physical and logical coordinates, namely $\mathbf{x}=\mathbf{x}(\mathbf{\xi})$. Note that this transformation depends on the monitor function $M$ which we can handle to produce a mesh with features appropriate for desired refinement criteria. For the change of coordinates, we notice that the covariant and contravariant vectors are given by

\begin{equation}
    \mathbf{a}_i=\frac{\partial \mathbf{x}}{\partial \xi_i},\quad \mathbf{a}^i=\nabla\xi_i,
\end{equation}

\noindent they are columns and rows of the Jacobian of $\mathcal{T}$ and $\mathcal{T}^{-1}$ as follows

\begin{equation}\label{eq:Jacobian}
    \mathcal{J}=[\mathbf{a}_1,\mathbf{a}_2,\mathbf{a}_3],\quad  \mathcal{J}^{-1}=\begin{bmatrix}
    (\mathbf{a}^1)^T\\
    (\mathbf{a}^2)^T\\
    (\mathbf{a}^3)^T
    \end{bmatrix}.
\end{equation}

\noindent With this in mind, the nabla operator in the physical domain, written in terms of the logical coordinates, is expressed in two possible forms

\begin{eqnarray}\label{eq:non-conservative}
\nabla &=& \sum_i \mathbf{a}^i\frac{\partial}{\partial\xi_i}\quad \quad\quad \text{{non-conservative}},\\\label{eq:conservative}
&=& \frac{1}{J}\sum_i \frac{\partial}{\partial \xi_i}J\mathbf{a}^i,\quad \text{{conservative}},
\end{eqnarray}

\noindent where $J=\text{det}\mathcal{J}$. The parabolic term in Eq. (\ref{eq:generalIVP}) involves second order derivatives and can be calculated using \eqref{eq:non-conservative} and \eqref{eq:conservative} as follows

\begin{equation}\label{eq:laplacian}
    \nabla\cdot (a\nabla u) = \frac{1}{J}\sum_{i,j}\frac{\partial}{\partial\xi_i} \left(a J \mathbf{a}^i\cdot \mathbf{a}^j\frac{\partial u}{\partial\xi_j}\right).
\end{equation}

\noindent 
With this term, we have the spatial part of Eq. (\ref{eq:generalIVP}) transformed into either the Physical or Logical domains.
 
The method becomes Adaptive-Moving if the transformation depends on time, in which case the coordinate transformation reads ${\bf x}={\bf x}({\bf \xi},t)$, and Eq. (\ref{eq:generalIVP}) has also to be modified as follows

 \begin{equation}\label{eq:unewcoordinates}
    u_t = \dot{u}-\nabla u\cdot\dot{x}.
 \end{equation}
 
\noindent A popular choice for $\dot{\mathbf{x}}$ is the so called {\it adaptive moving mesh partial differential equation} defined as the time derivative of the adaptation functionals \eqref{eq:func1} or \eqref{eq:func2} \cite{Huang2001}. For the case we are considering, \eqref{eq:func2} results in the time derivative of $\mathbf{x}$ equals to the Adaptive Mesh Equation \eqref{eq:ELEQ} 

\begin{equation}\label{eq:MMPDE5}
        \dot{\mathbf{x}} = \frac{1}{\tau}\nabla_\xi\cdot\left(\rho \nabla_\xi \mathbf{x}\right),
\end{equation}

\noindent where the constant $\tau$ controls the mesh speed of response. With equations \eqref{eq:laplacian} and \eqref{eq:unewcoordinates} it is possible to rewrite the expressions for the Schr\"odinger equation for a general change of coordinates, solution of equations \eqref{eq:ELEQ} and \eqref{eq:MMPDE5}. This completes the transformation of the general Eq. (\ref{eq:generalIVP}).

\subsection{Mesh refinement generator and the monitor function}

\begin{figure}
\includegraphics[width=4cm]{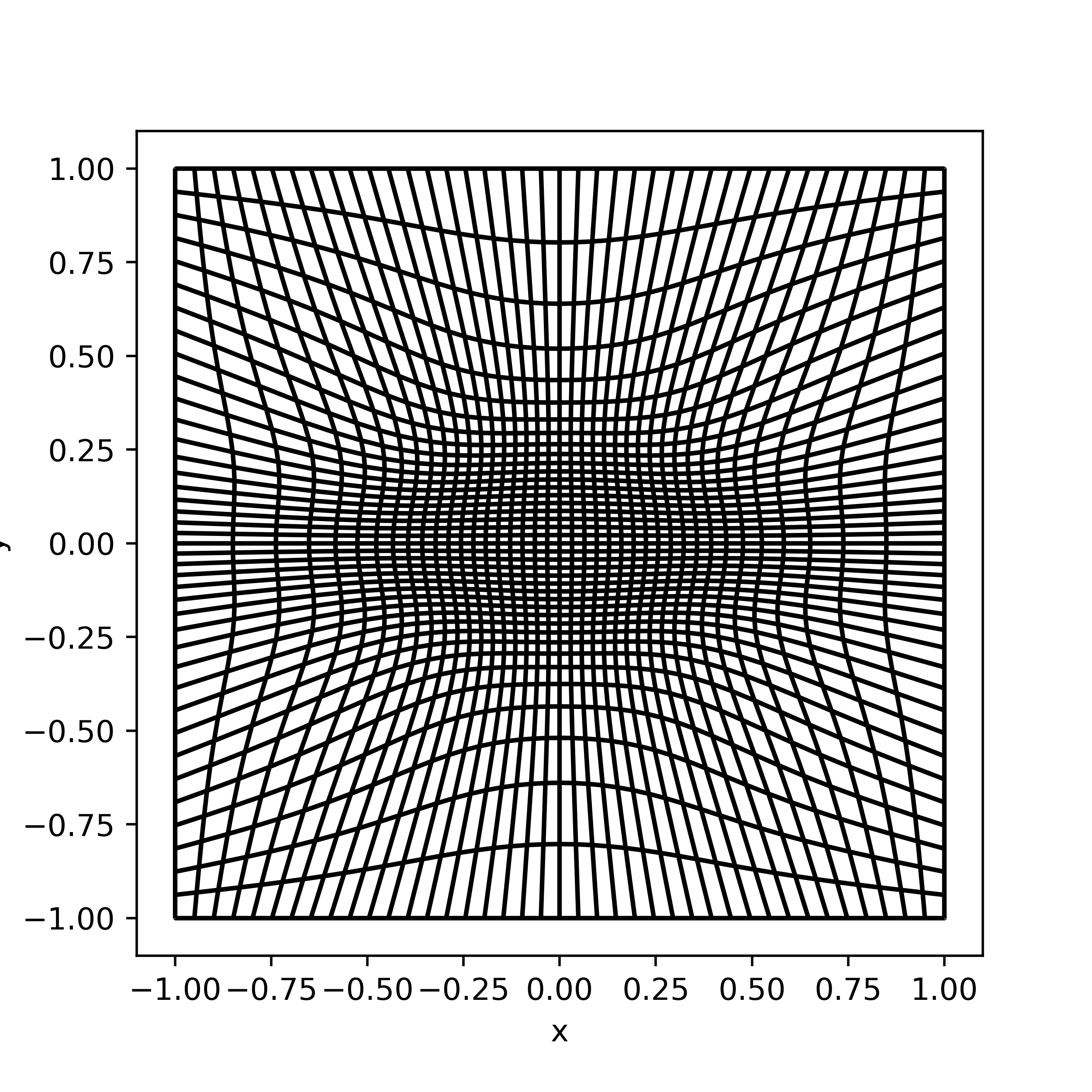}\includegraphics[width=4cm]{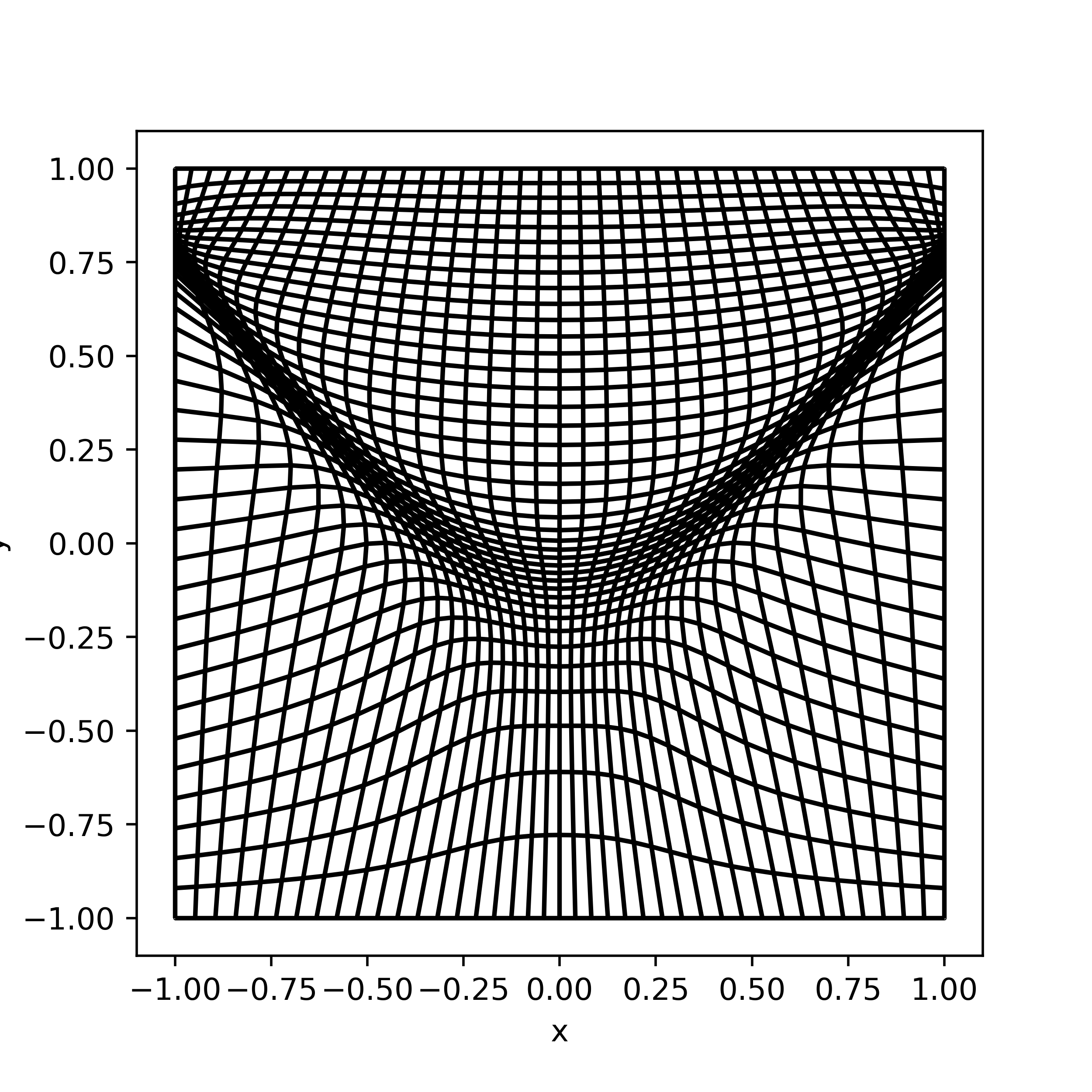}
\includegraphics[width=4cm]{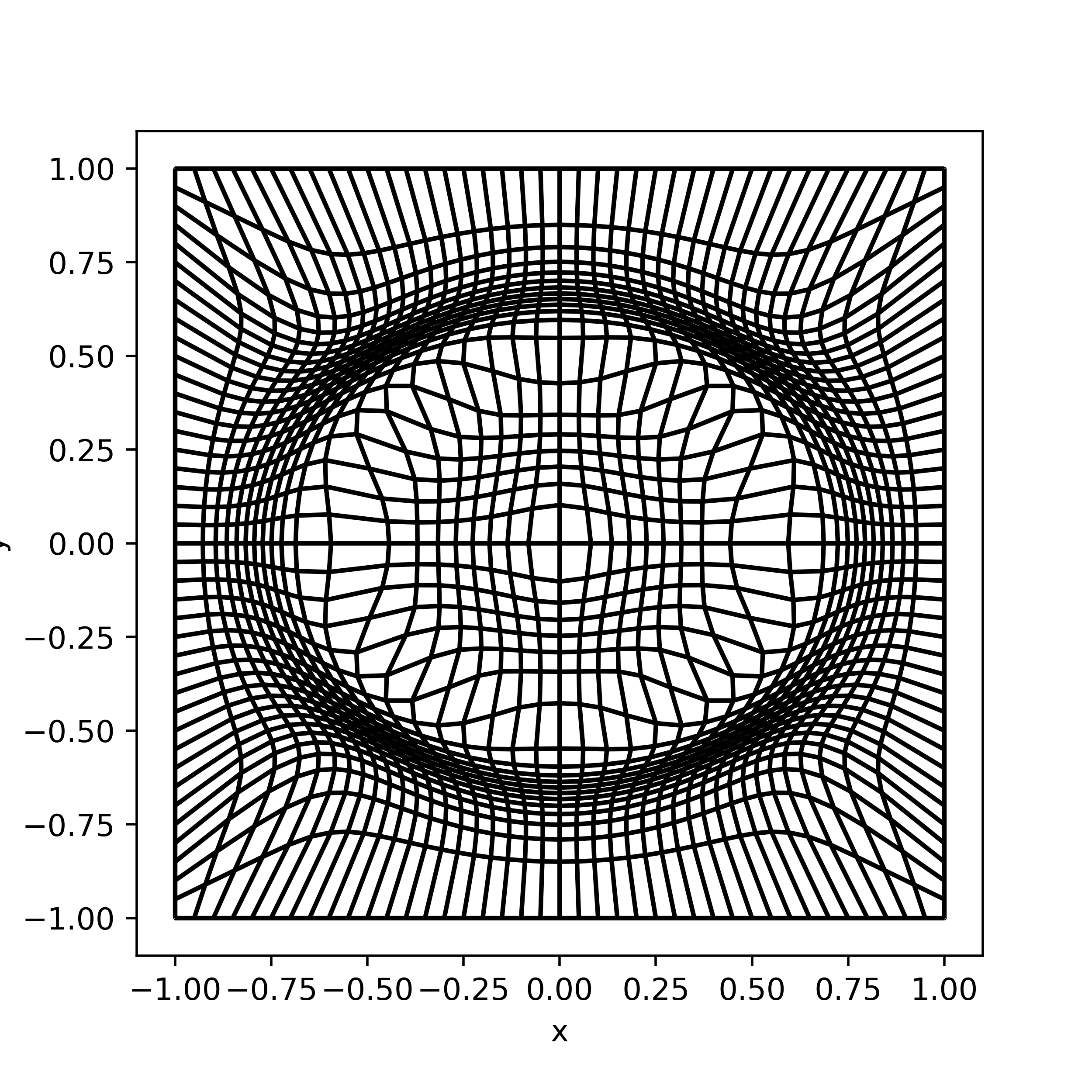}\includegraphics[width=4cm]{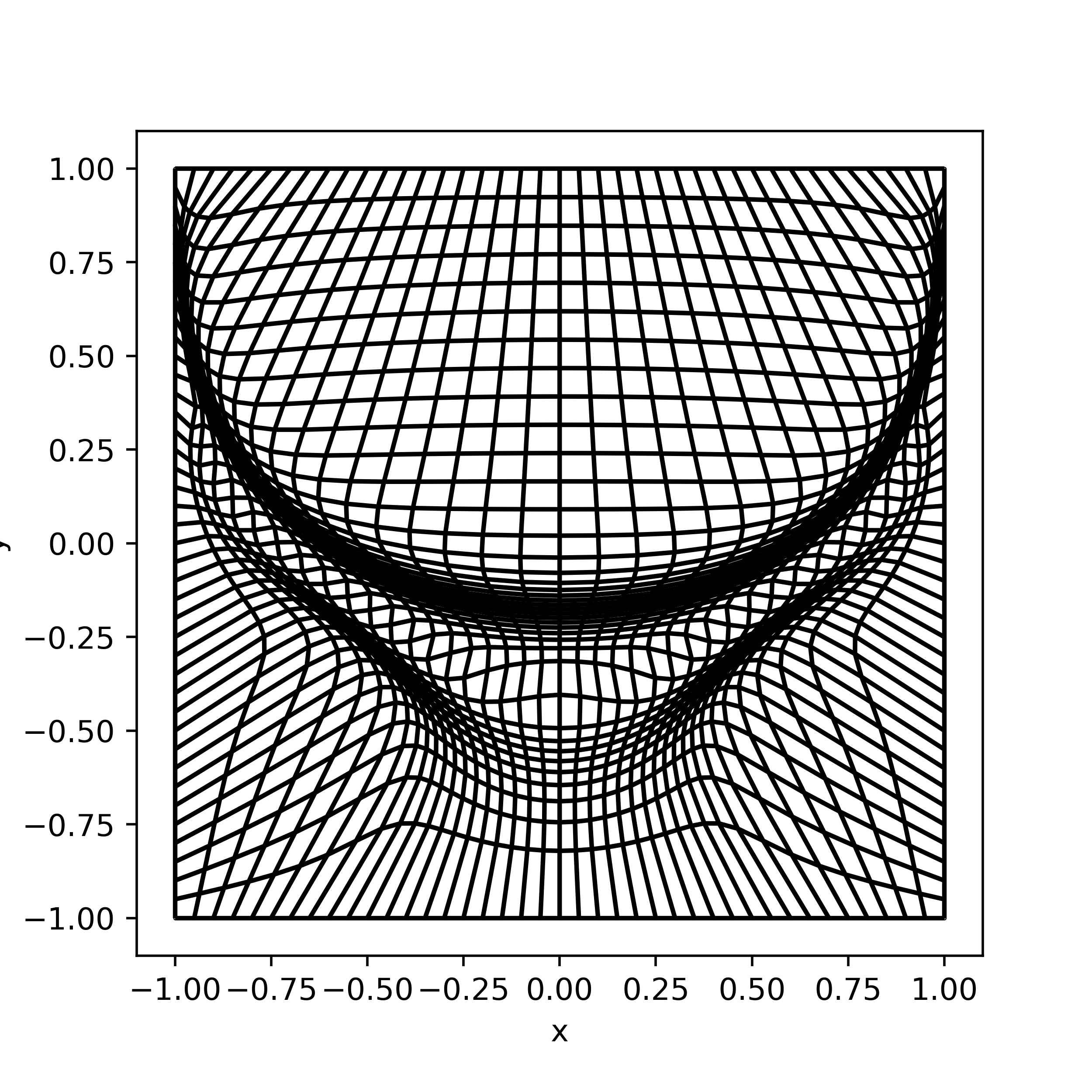}
\caption{Physical mesh in the $\kappa = 0$ plane for the four combinations of $\omega_1,\omega_2,v_1,v_2$. At the top we show the two combinations for $\omega_1$, at the bottom the cases for $\omega_2$, at the left column the cases $v_1$ and at the right those for $v_2$. For these meshes we fix the parameter $\alpha=100$.}\label{fig:Test1}
\end{figure}

The set of equations \eqref{eq:ELEQ} determines the connection between logical and physical domains. This expression is known as the {\it Adaptive Mesh Equation} and depends on the mesh density $\rho = |\omega|^{3/2}$. This function can take different forms and the choice depends on the Partial Differential Equation (PDE) to be solved. For illustration we consider two possible expressions for $\omega$:

\begin{eqnarray}\label{eq:rho1}
\omega &=&\omega_1 =\sqrt{1+\alpha v^2},\\\label{eq:rho2}
\omega &=&\omega_2 =\sqrt{1+\alpha |\nabla_\xi v|^2},
\end{eqnarray}

\noindent where $\alpha$ parameter regulates the mesh ``stress" through the function $v$. Again for illustration, we consider two different expressions of the resolution concentration function $v$: 

\begin{eqnarray}\label{eq:ex1}
v(x,y,z) &=& v_1 =\text{e}^{-(5x^2+7y^2-1)^2/10},\\\label{eq:ex2}
v(x,y,z) &=&v_2=\text{e}^{-10(y-x^2+0.5)^2},
\end{eqnarray}

\noindent needed in \eqref{eq:rho1} and \eqref{eq:rho2}. The physical domain mesh obtained by using the four combinations of $\omega_1,\omega_2,v_1,v_2$ is shown in Figure \ref{fig:Test1}, which actually are standard tests of the AMM method \cite{HuangAMM,Tang:2006,Huang:2006,Huang2001,Weiming:2006}.
At the top, we show the case for $\omega_1$ and the two expressions $v_1$ and $v_2$, which shows a higher physical resolution where $v^2$ is around its maximum. At the bottom we show the results for $\omega_2$ and $v_1,v_2$; in this case, the higher resolution for the physical mesh is concentrated in the region where $|\nabla_\xi v|^2$ is maximum. 

A particular setup useful to solve the SP system, with some initial conditions, uses functions $v$ and $\omega$ that concentrate a nearly constant high resolution at the center of the domain and a constant coarse resolution in the outskirts. A function $v$ that helps this purpose, is the following

\begin{equation}\label{eq:funct_sp}
    v=A\left(2+\tanh\left(\frac{R-r_c}{\delta}\right)-\tanh\left(\frac{r_c}{\delta}\right)\right).
\end{equation}

\begin{figure}
\includegraphics[width=4cm]{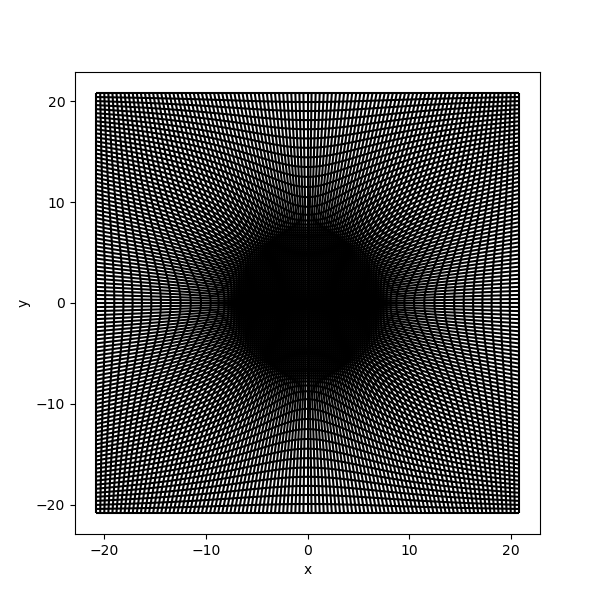}
\includegraphics[width=4cm]{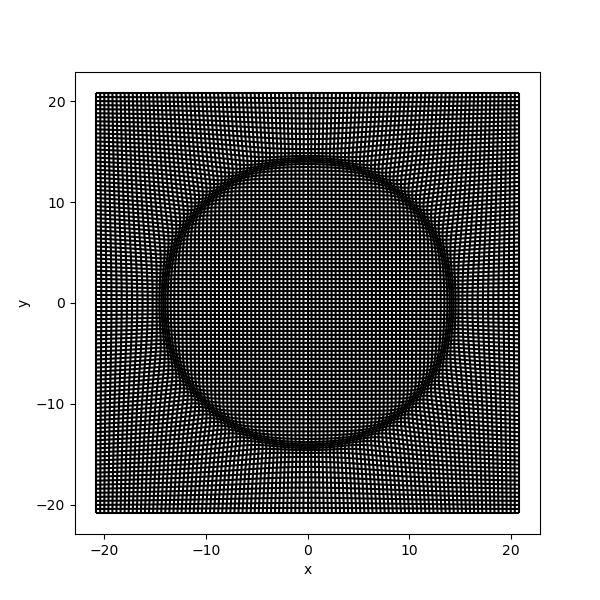}
\caption{Example of a physical domain $[-20.8,20.8]^3$ discretized using $N=104^3$ cells, projected on the $\kappa = 0$ plane $z=0$. In this specific case we use $v$ given by Eq. (\ref{eq:funct_sp}) with parameters $R=\sqrt{x^2+y^2+z^2}$, $A=2.7$, $r_c=15$ and $\delta=1$. At the left we show the result for $\omega=\omega_1$ which defines a region with high resolution $\sim 0.2$ in the center and a coarse resolution near the boundary of $\sim 0.4$. At the right we show the result for $\omega=\omega_2$ that defines a spherical shell of high resolution. From these two cases we use $\omega=\omega_1$ in various simulations below.} 
\label{fig:BoostedAMM}
\end{figure}

\noindent The result for $\omega=\omega_1,~\omega_2$, $R=\sqrt{x^2+y^2+z^2}$, $A=2.7$, $r_c=15$, $\delta=1$, on the physical domain $[-20.8,20.8]^3$, discretized with $N_{\xi}=N_{\eta}=N_{\kappa}=104$ cells along each direction, is shown in Figure \ref{fig:BoostedAMM}. For this example we actually consider the domain useful for the evolution of interesting scenarios for the SP system and resolutions that will be used in some examples below. Notice that $\omega_1$ provides the needed mesh with a high-density central resolution of $\sim 0.2$ and a coarse resolution of $\sim 0.4$ in the outskirts. Notice also that $\omega_2$ concentrates resolution on a spherical ring and maintains the coarse resolution everywhere else. 

In summary, the implementation of the AMM method can be described in a number of simple steps. 
For this, consider there is a method to evolve the solution $u^n$ from time $t^n$ to the solution $u^{n+1}$ at time $t^{n+1}$ in a uniformly discretized domain, in our case this role is played the logical domain $\Omega_L$. The steps needed to construct this evolution on the logical domain are the following:

\begin{enumerate}
    \item Consider the objective is to construct a numerical solution of equation (\ref{eq:generalIVP}) in the numerical physical domain $\Omega_P$ along with the discrete-time levels as $t^n$.
    \item At time $t^n$, construct a version of equation (\ref{eq:generalIVP}) for $u(\xi_i,\eta_j,\kappa_k,t)$ described in coordinates $(\xi_i,\eta_j,\kappa_k)\in\Omega_L$ using the following steps:
        \begin{itemize}
            \item[2.1] Define functions $\omega$ and $v$, for example those in (\ref{eq:rho1},\ref{eq:rho2}) and (\ref{eq:ex1}, \ref{eq:ex2}) respectively.
            \item[2.2] With this information, solve equation (\ref{eq:ELEQ}) and construct transformations $\mathcal{T}$ and $\mathcal{T}^{-1}$.
            \item[2.3] By knowing the transformations it is possible to calculate the Jacobian of the transformations $\mathcal{J}$ and $\mathcal{J}^{-1}$ in (\ref{eq:Jacobian}). Then use the chain rule to construct space derivative operators in the logical domain through expressions (\ref{eq:non-conservative},\ref{eq:conservative},\ref{eq:laplacian}).
            \item[2.4] It also allows one to use the chain rule for the construction of the time derivative of $u(\xi,\eta,\kappa)$ through Equation (\ref{eq:MMPDE5}).
        \end{itemize}
    \item Use an evolution method within the Logical domain $\Omega_L$ to evolve the solution and the physical coordinates from $u^n(\xi_i,\eta_j,\kappa_k)$ to $u^{n+1}(\xi_i,\eta_j,\kappa_k)$ and $x_i^n(\xi_i,\eta_j,\kappa_k)$ to $x_i^{n+1}(\xi_i,\eta_j,\kappa_k)$ respectively.
    \item Finally, the solution in the physical domain is calculated by transforming back the solution from $\Omega_L$ to $\Omega_P$ with $u^{n+1}(x_i,y_j,z_k)=u^{n+1}(\mathcal{T}(\xi_i,\eta_j,\kappa_k))$.
\end{enumerate}

This set of steps is applied from initial time $n=0$ until a desired number of time steps, using the evolution scheme described in the next section.

\section{Application to the Schr\"odinger equation}
\label{sec:GPP}

\subsection{Initial Conditions}

We consider various scenarios involving the Schr\"odinger equation. Problem A corresponds to a particle on a harmonic trap, where the initial conditions used corresponds to the exact solution for a given number of nodes. Problems B correspond to various cases involving the SP system of equations, that start with stationary solutions of the SP system, where the wave function for the stationary solution is injected in the numerical domain using interpolation; in this case, after the wave function is injected, consistent initial data require the solution of Poisson equation at the initial time. Finally, Problem C corresponds to the evolution of a toy problem of a dark matter subhalo orbiting a host halo with a fixed density profile and gravitational potential. In this case, the wave function of the subhalo is again an equilibrium configuration at the initial time and is evolved in the test field regime in order to see its disruption process.

\subsection{Evolution}

The evolution of the various examples is ruled by Schr\"odinger equation, and in the case of the SP system, its potential has to fulfill the Poisson equation during evolution.

In order to solve Schr\"odinger equation we use the second-order accurate implicit Crank-Nicholson finite differences scheme \cite{NumericalRecipes}. The method starts by writing Schr\"odinger equation at the points of the logical domain $(\xi_i,\eta_j,\kappa_k,t^n)$ and $(\xi_i,\eta_j,\kappa_k,t^{n+1})$, where the wave function evaluated at each time step is written as $\psi^{n}_{i,j,k}$ and $\psi^{n+1}_{i,j,k}$ respectively and these values are related as follows 

\begin{equation}\label{CN}
\left(\mathbb{I}+\mathbb{M}\right)\psi^{n+1}_{i,j,k}=\left(\mathbb{I}-\mathbb{M}\right)\psi^n_{i,j,k},
\end{equation}

\noindent where $\Delta t$ is time resolution, $\mathbb{M}=\frac{i}{2}\hat{H}\Delta t$ and $H$ is the Hamiltonian discrete operator defined on the logical domain $\Omega_L$. 
Equation (\ref{CN}) is a linear system of equations whose unknowns are the values of the wave function $\psi^{n+1}_{i,j,k}$ at time $t^{n+1}$.

This evolution scheme suffices to solve Problem A (\ref{sec:TestA}), where the potential term in the Schr\"odinger equation is a fixed function of the spatial coordinates. For Problems B (\ref{sec:tests_SP}) and C (\ref{sec:ProbC}), it is still necessary to solve the Poisson equation during the evolution for the gravitational potential that enters back into Schr\"odinger equation (specifically the subhalo potential for Problem C).

Poisson equation, being elliptic, is solved on the discrete logical domain using the Multigrid method with a three resolution levels V cycle at each time step $t^n$. \cite{NumericalRecipes}.

\section{Specific Problems}
\label{sec:specificproblems}

In this Section we solve three problems in order to test the AMM approach. For comparison, we use a code that works using Fixed Mesh Refinement \cite{Guzman2014} in Problems A and B. Problem C is related to a more dynamical scenario and is solved only with the AMM code. In this case, the AMM code shows the ability to track the evolution of deformed density profiles and can be related to astrophysical scenarios.

\subsection{Problem A: Schr\"odinger Equation for a particle in a Harmonic Oscillator potential}
\label{sec:TestA}
We aim to test the implementation of the Schr\"odinger equation for a non-trivial case of an exact solution, which allows one to assess the numerical results. 
This basic Test was carried out using the AMM off-mode 
in order to test the Schr\"odinger equation integrator exclusively in a very simple situation where the density $|\psi|^2$ remains concentrated in a fixed region. In this case Schr\"odinger equation reads 

\begin{equation}
\hat{H}\Psi=-i\hbar\frac{\partial\Psi}{\partial t} = -\frac{\hbar^2}{2m}\nabla^2\Psi + \frac{1}{2} m\omega^2 (x^2+y^2+z^2)\Psi,
\end{equation}

\noindent where $\hat{H}$ is the Hamiltonian operator corresponding to a tridimensional harmonic oscillator.

Initial conditions correspond to the solution of the stationary exact solution $\psi_s$ of the above equation which obeys 
    $\frac{E}{\hbar\omega}\psi_s = -\frac{\hbar^2}{2m}\nabla^2\psi_s + \frac{1}{2} m\omega^2 (x^2+y^2+z^2)\psi_s,$
where $\hat{E}=(n_x+n_y+n_z+\frac{3}{2})\hbar\omega$, whose solution is given by: 

\begin{eqnarray}
\psi_s(x,y,z) &=& C(n_x,n_y,n_z)H_{n_x}(x)H_{n_y}(y)H_{n_z}(z)\nonumber\\\label{eq:HarmExact}
&&\times \text{e}^{-\beta^2\frac{x^2+y^2+z^2}{2}},
\end{eqnarray}

\noindent where $H_{n_i}(x_i)$ are the Hermite polynomials of order $n_i$ and the normalization constant is given by

\begin{eqnarray}
    C(n_x,n_y,n_z)&=&\left(\frac{\beta^2}{\pi}\right)^{3/4}\frac{1}{\sqrt{2^{n_x+n_y+n_z}n_x!n_y!n_z!}},\nonumber\\
    \beta &=& \sqrt{\frac{m\omega}{\hbar}}.\nonumber
\end{eqnarray}

Problem A consists in the evolution of this initial condition for the particular state $n_x=n_y=n_z=2$, and $m=\omega=\hbar=1$, and it is expected that the evolution reproduces the properties and behavior of the fully time dependent wave function $\Psi(x,y,z,t)=e^{-iEt/\hbar}\psi_s(x,y,z)$.

The numerical parameters used are the following. The domain is the box $(x,y,z)\in [-10,10]\times[-10,10]\times[-10,10]$, discretized with resolution $\Delta x=\Delta y=\Delta z=0.2$.

\begin{figure*}[h]
\includegraphics[width=7cm]{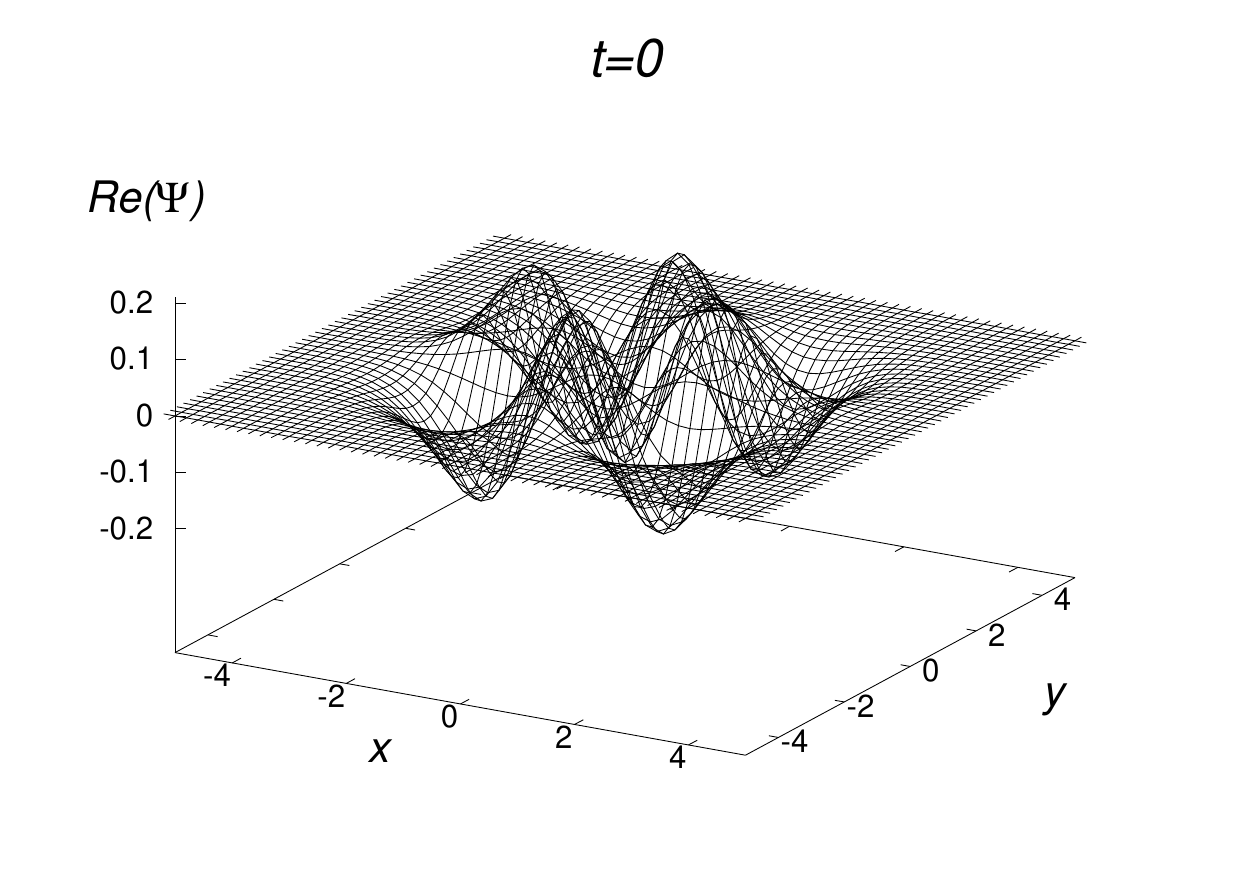}
\includegraphics[width=7cm]{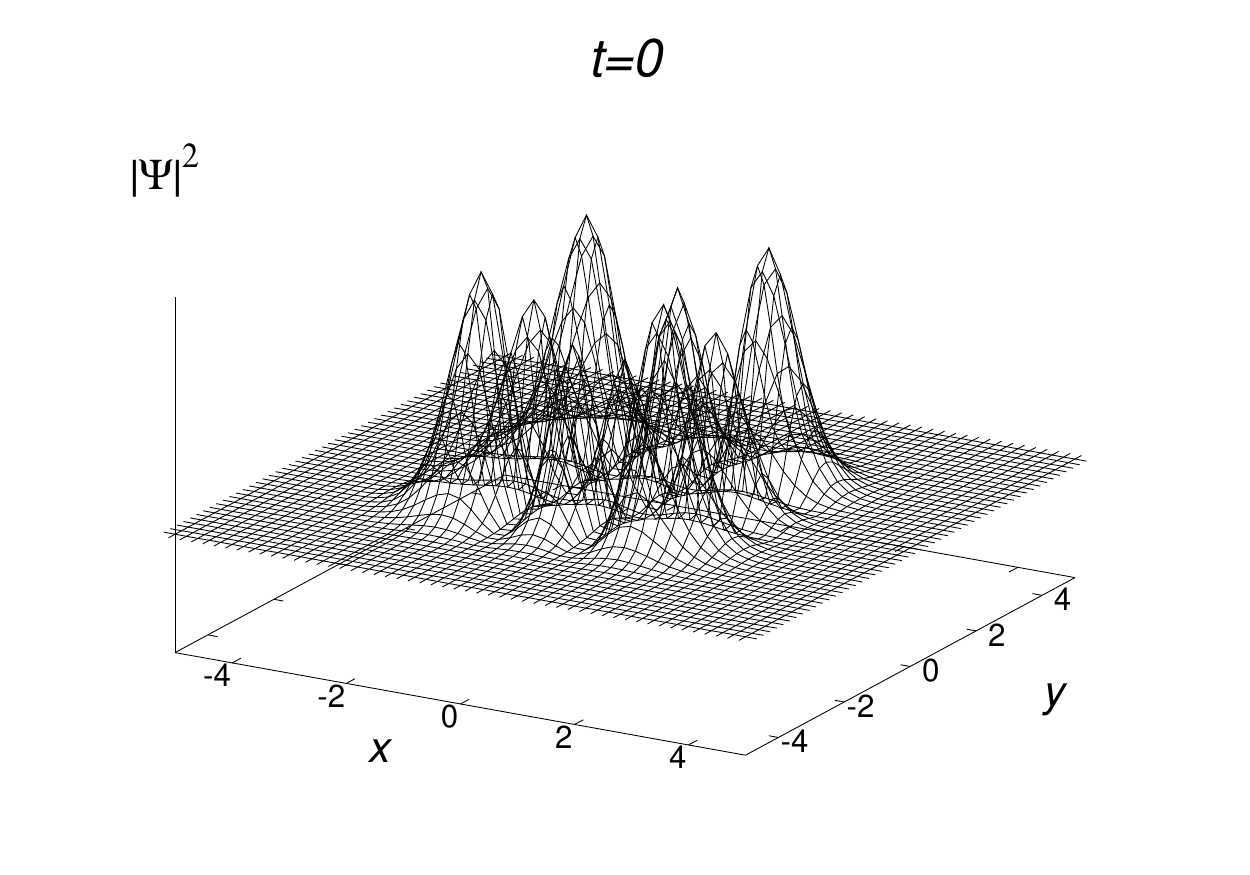}
\includegraphics[width=7cm]{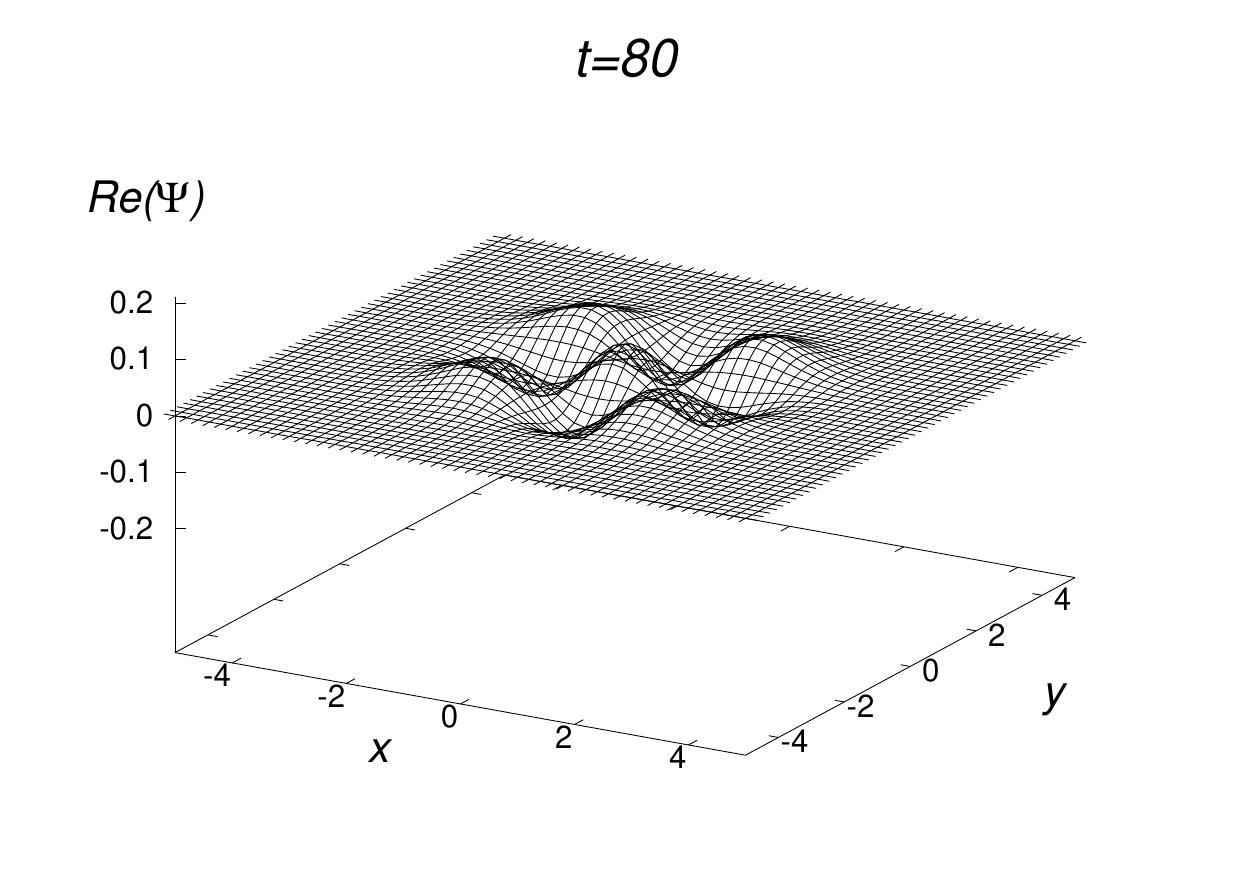}
\includegraphics[width=7cm]{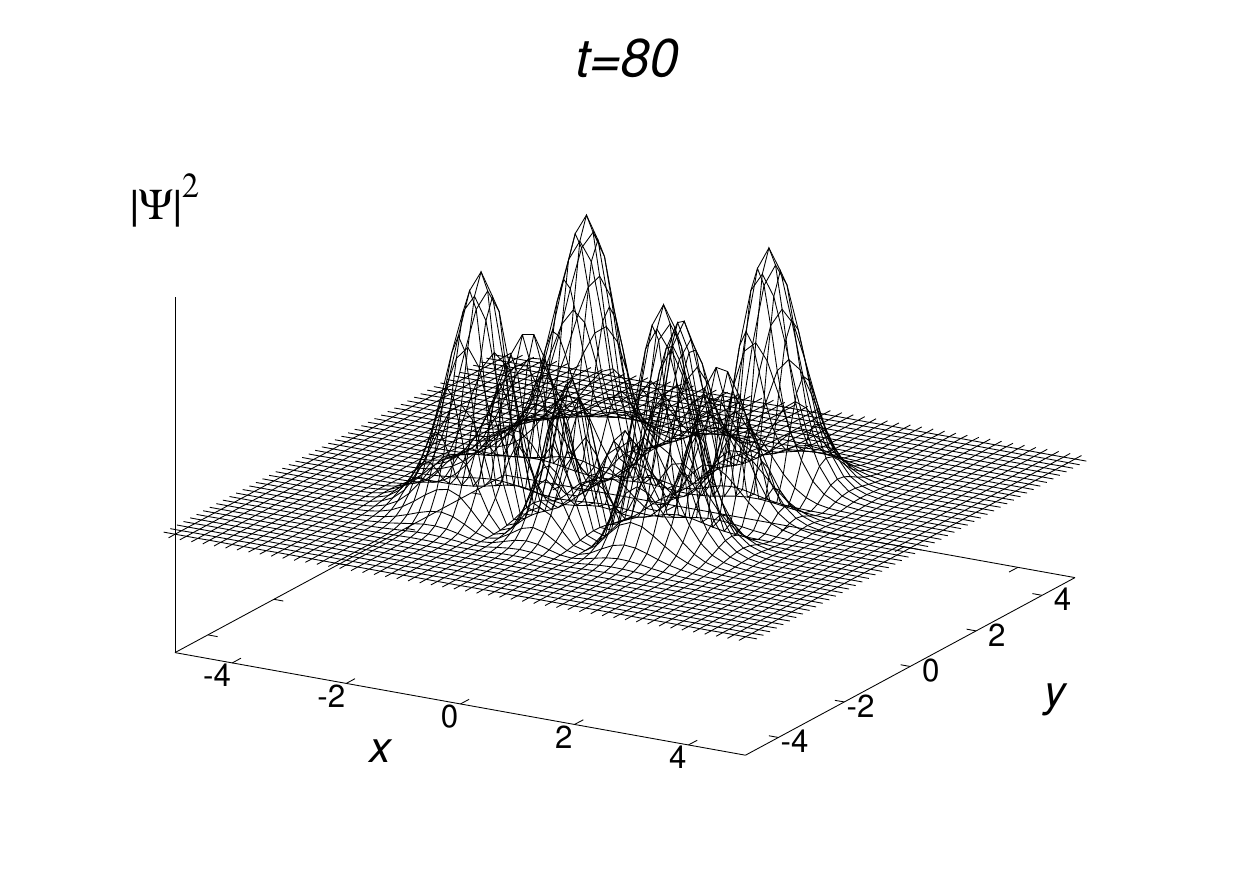}
\includegraphics[width=7cm]{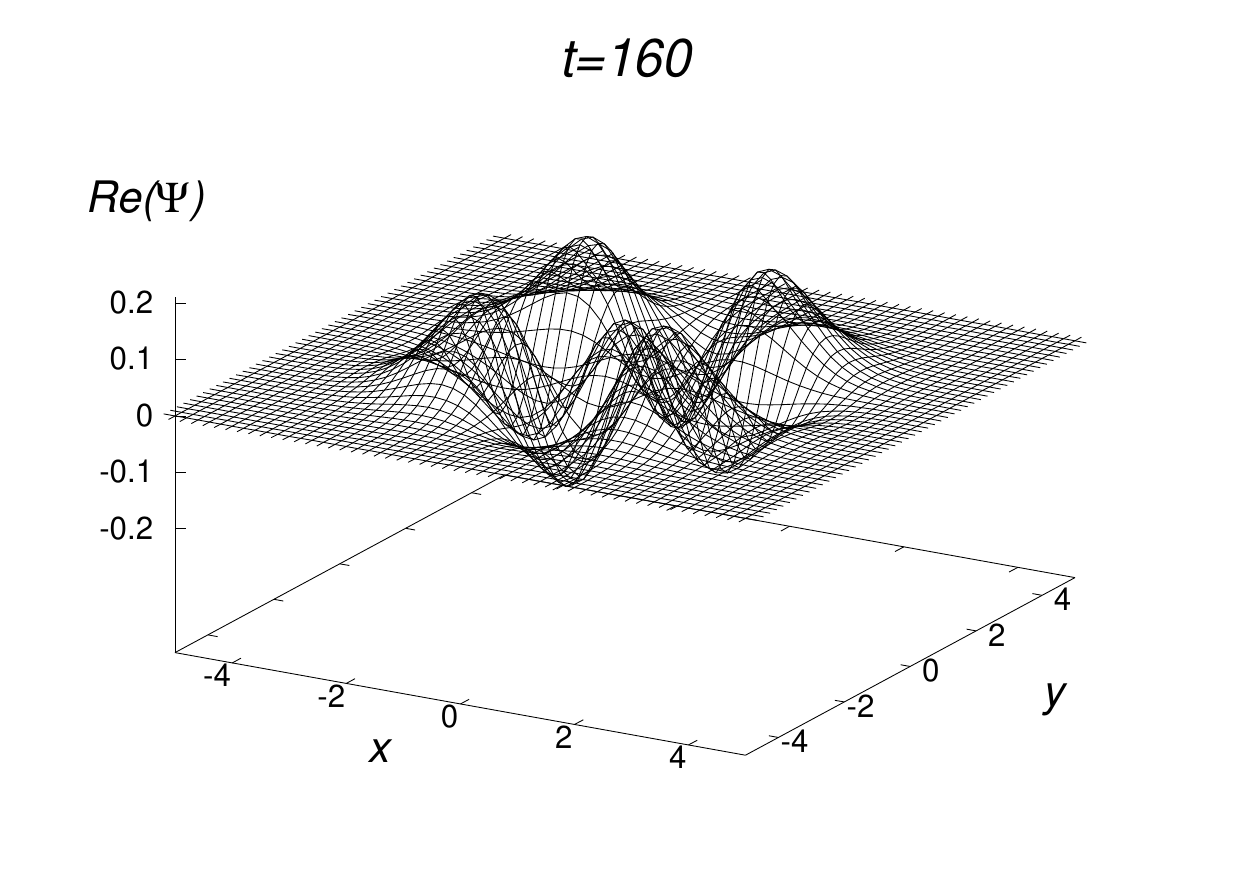}
\includegraphics[width=7cm]{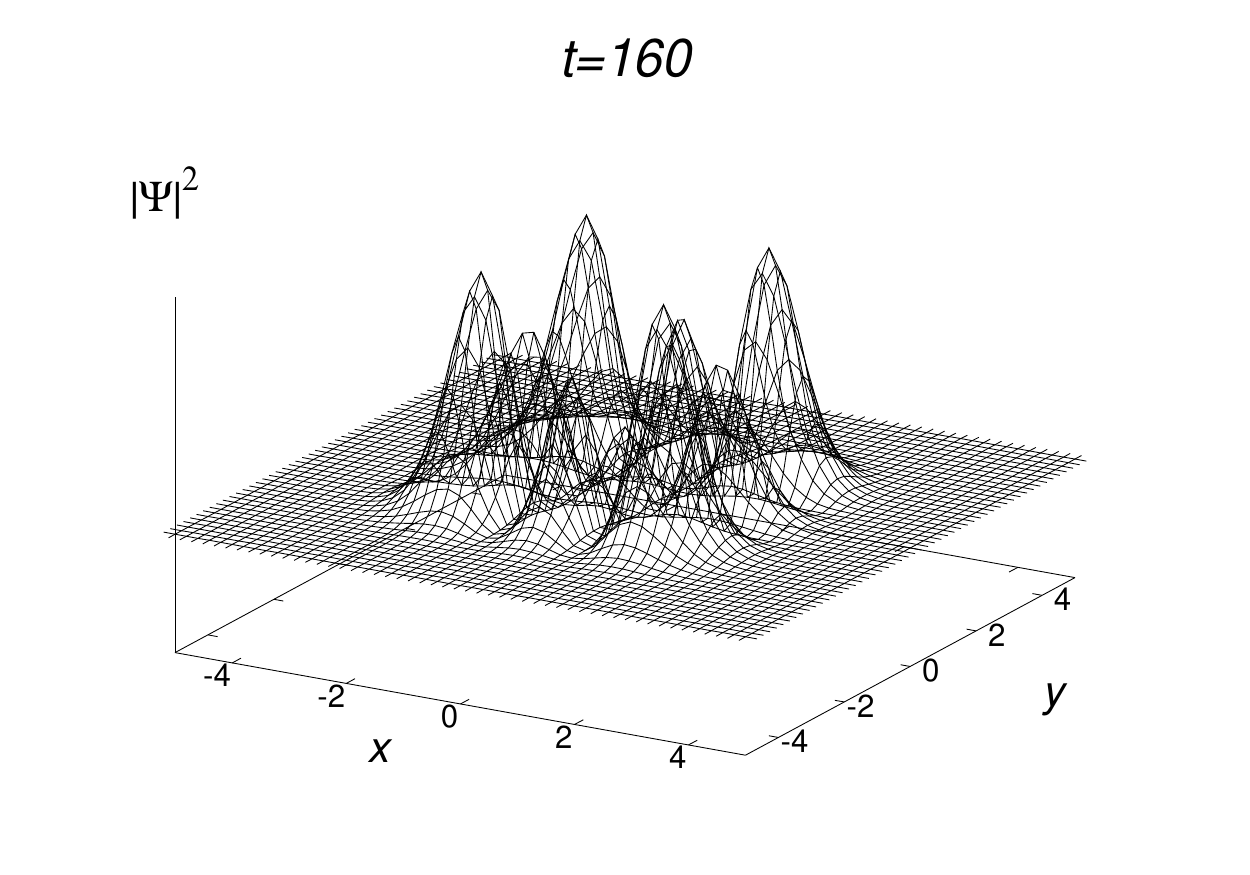}
\caption{Snapshots of the projection of Re($\Psi$) and $|\Psi|^2$ on the $xy-$plane. This illustrates that the wave function is evolving whereas the density remains nearly stationary.} 
\label{fig:snaps}
\end{figure*}

\begin{figure*}[h]
\includegraphics[width=7cm]{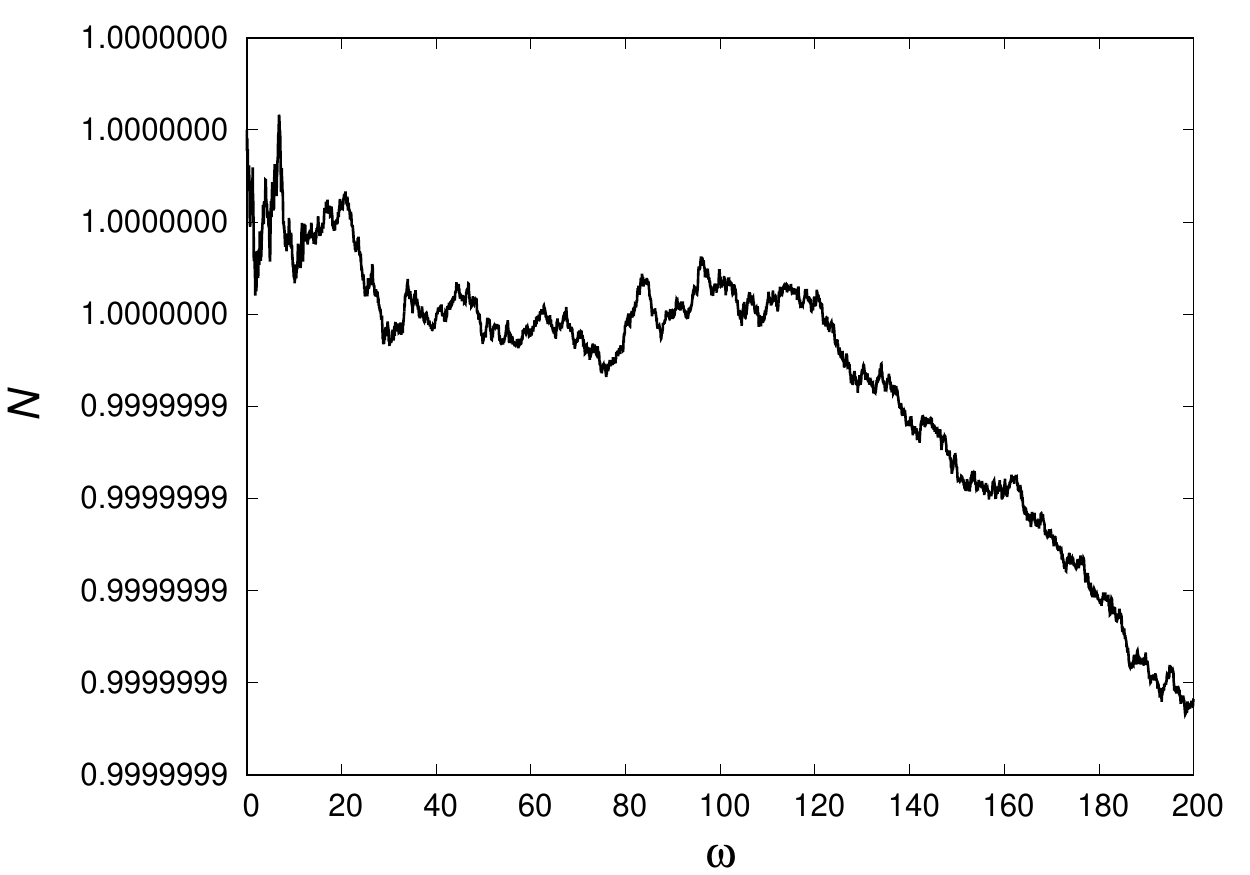}
\includegraphics[width=7cm]{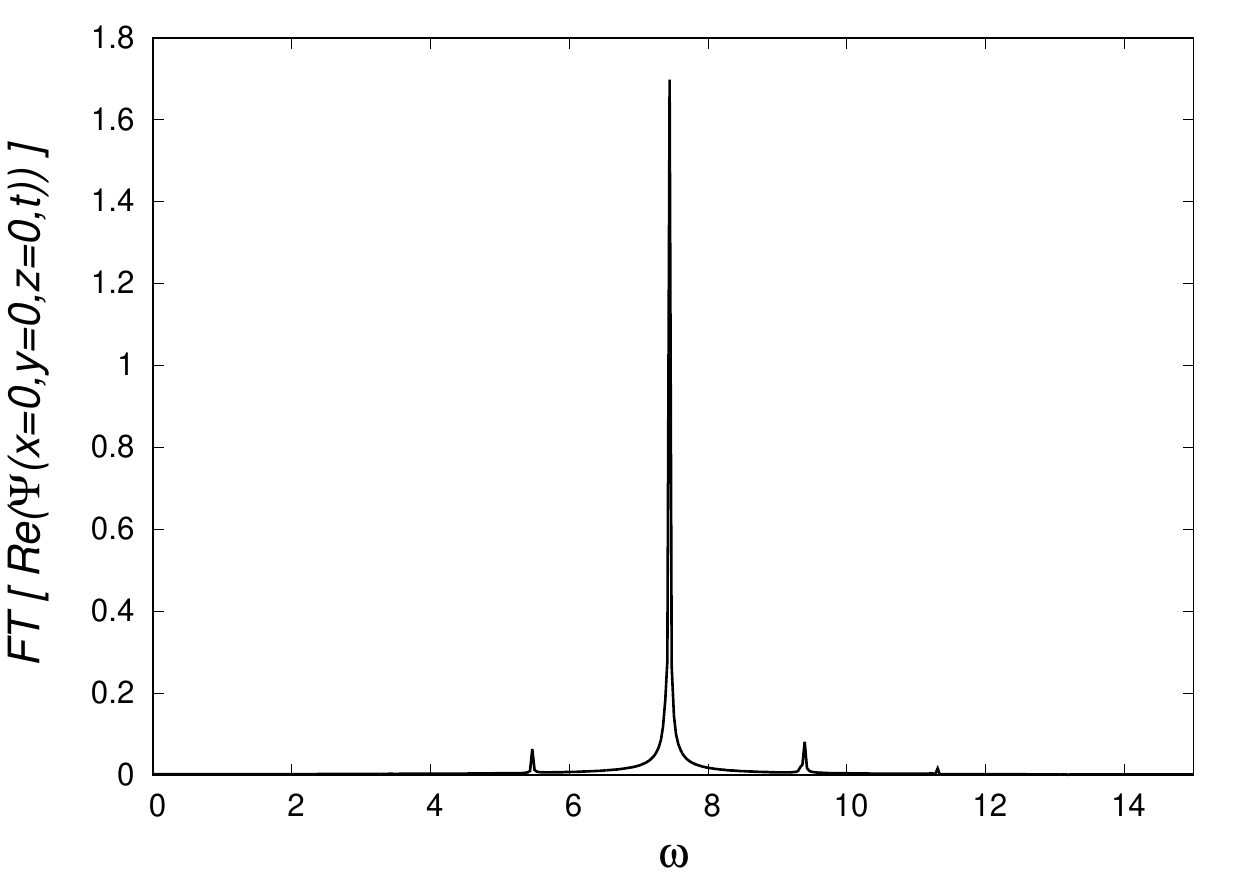}
\includegraphics[width=7cm]{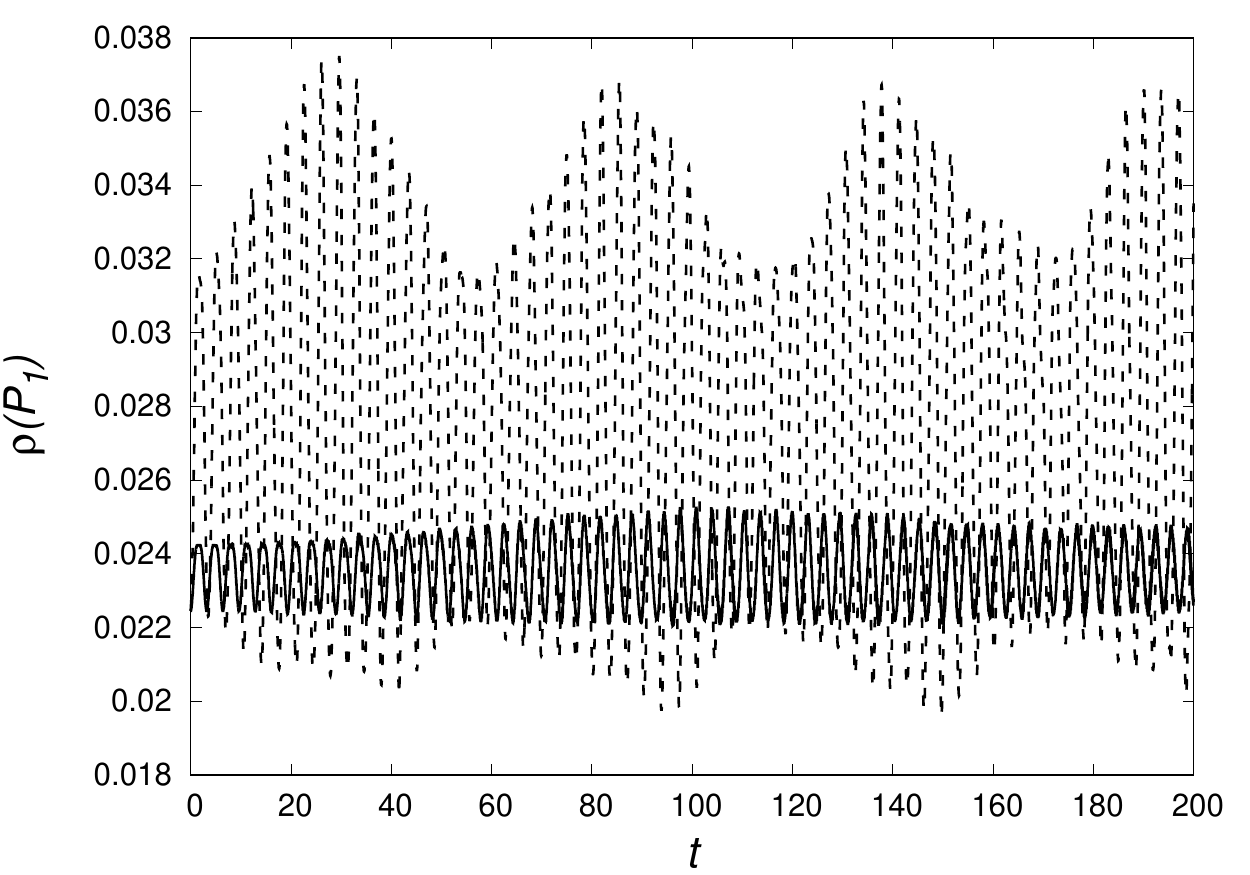}
\includegraphics[width=7cm]{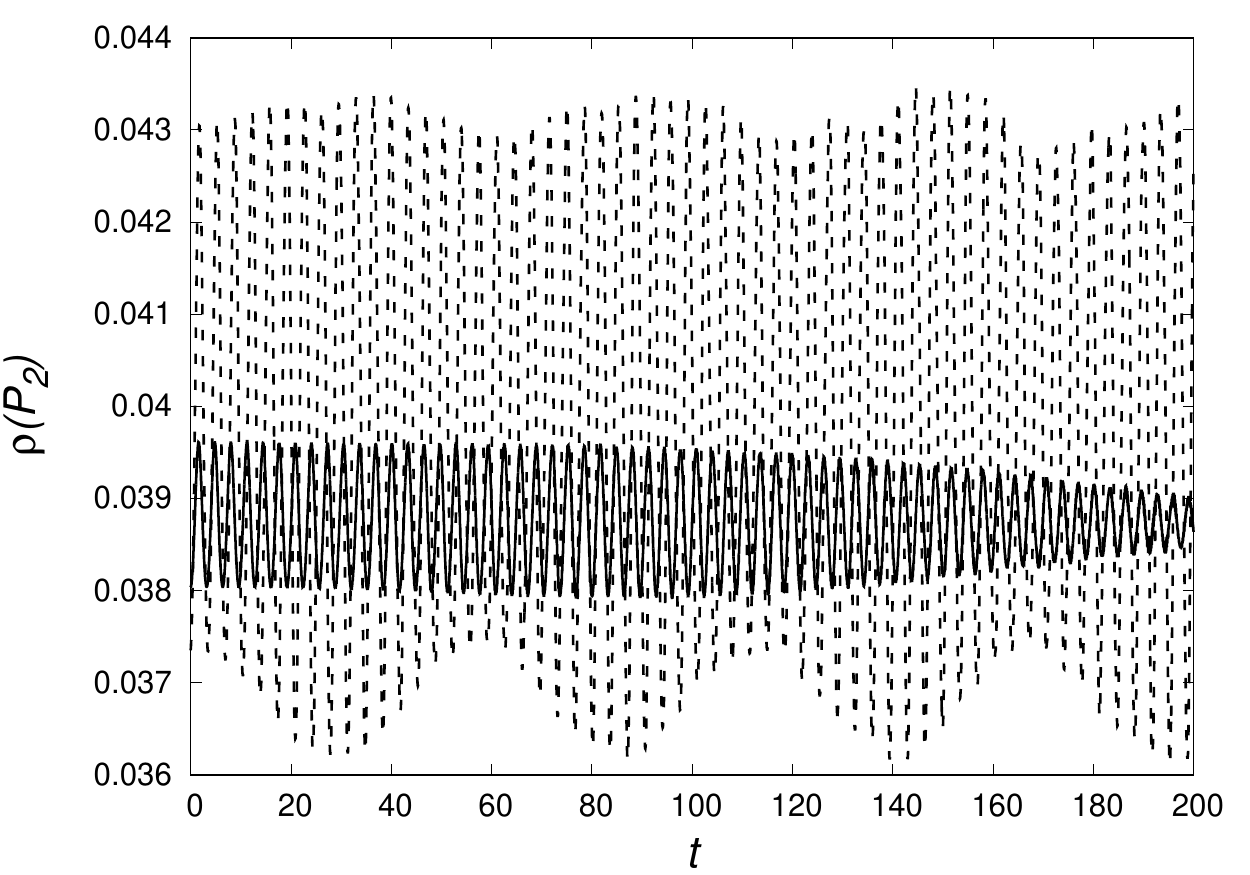}
\caption{(Top-left) Time-series corresponding to the total number of particles within the domain given by $N=\int |\Psi|^2 d^3 x$, this function remains close to one during the evolution. (Top-right) Fourier Transform of the central value of the real part of the wave function. The main peak is located precisely at the frequency $(2+2+2+3/2)$ which coincides with that of the exact solution. { (Bottom) Convergence of the numerical solution at representative local maximum points $P_1=(0,0,0)$ and $P_2=(\sqrt{2.5},\sqrt{2.5},0)$ toward the exact solution, using the two resolutions $\Delta xyz =0.4$ and $\Delta xyz =0.2$}. } 
\label{fig:Repsi}
\end{figure*}

We show the numerical solution is stationary by checking the density remains nearly time-independent, whereas the wave function oscillates with the frequency $\omega$. In Fig. \ref{fig:snaps} some snapshots of $Re(\Psi)$ and $|\Psi|^2$ are shown at various times, illustrating how the wave function evolves and the density remains nearly time independent. In Fig. \ref{fig:Repsi} we show that $N=\int |\Psi|^2d^3x$, the number of particles within the numerical domain, remains close to one by less than one part in a million during 250 oscillations of the wave function, which indicates that the evolution is nearly unitary.

In order to assess in more detail the numerical solution, we compare the oscillation frequency of $Re(\Psi)$ of a numerical solution with the frequency of the exact solution. For this, we compute a Fourier transform (FT) of the central value of $Re(\Psi)$. As shown in Figure \ref{fig:Repsi} the FT shows a well-defined peak at the theoretical value $(2+2+2+3/2)$. In summary, all these results together indicate that the evolution solves correctly the Schr\"odinger equation in a non-trivial but well-known case. 
{ Additionally, in order to provide a convergence test we calculate interpolated values of the density a the local maximae $P_1(0,0,0)$ and $P_2(\sqrt{2.5},\sqrt{2.5},0)$ as a function of time using two resolutions $\Delta xyz =0.4$ and $\Delta xyz =0.2$, and show the results also in Figure \ref{fig:Repsi}, where convergence toward the exact values $.02245$ for $P_1$ and $0.0387$ is manifest.}

\subsection{Problem B: the Schr\"odinger-Poisson System}
\label{sec:tests_SP}

In this section we use the AMM method to solve the SP system, and since there are no exact analytic solutions, we compare the solutions with those constructed using the xBEC code, that implements the FMR method \cite{xBEC}. This code is set to solve Sch\"odinger equation in time with the Crank-Nicholson and Alternating Direction Implicit (ADI) approaches, in order to have similar integration methods.
Each particular case of the SP problem involves the evolution of a specific set of initially equilibrium configurations of the Schr\"odinger-Poisson system solved as described in \cite{GuzmanUrena2004}.

For the three problems in B, we use the same numerical domain $[-20,20]^3$ with a base resolution $\Delta xyz =0.4$. Unigrid mode of xBEC uses a simple discretization of the domain $[-20,20]^3$ with resolution $\Delta x=\Delta y=\Delta z =0.4$. The FMR mode uses the same numerical domain, but this time with the additional subdomain $[-10,10]^3$ discretized with resolution $\Delta x=\Delta y=\Delta z =0.2$, which will increase the accuracy of the solution within this box.

On the other hand, for the AMM code, {\it mode off} sets the logical and the physical meshes as uniform discretizations with resolution $\Delta x=\Delta y=\Delta z =0.4$. The {\it mode on} considers a transformation that on one hand defines a mesh with coarse resolution $\Delta x=\Delta y=\Delta z =0.4$ where gradients of density are small and, on the other hand, nearly double resolution $\Delta x\Delta y=\Delta z =0.2$ in the regions where density gradients are high. The idea is that the numerical setup with xBEC and AMM are similar in resolution and accuracy.

\subsubsection{Problem B.1: evolution of an equilibrium configuration}
\label{sec:TestB1}

As a first problem, we consider the evolution of a ground state stationary spherical configuration. The numerical domain is set to the box $[-20,20]^3$.
For comparison, both the control code xBEC is used in unigrid/FMR mode, and for comparison, the AMM code is set in the equivalent off/on mode.

\begin{figure}
\includegraphics[width=7cm]{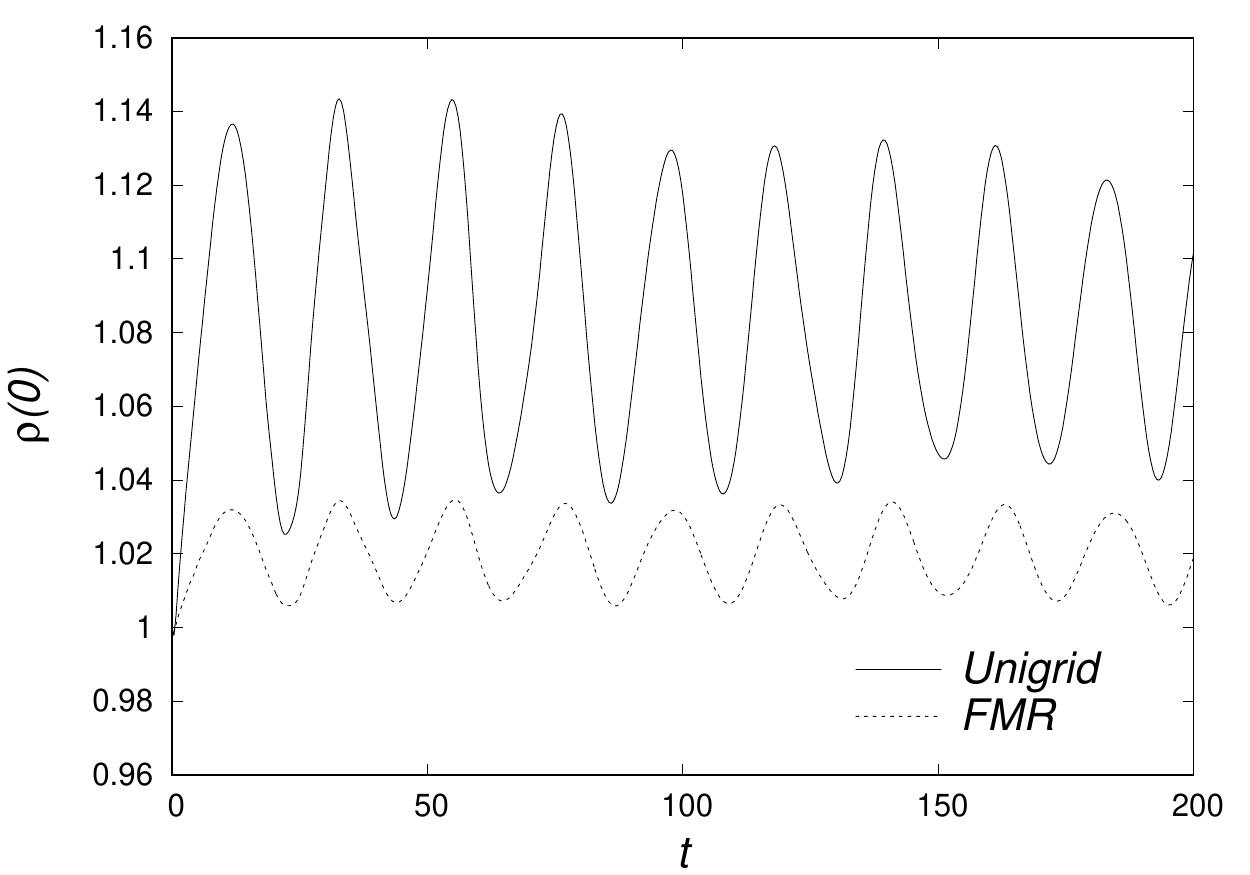}
\includegraphics[width=7cm]{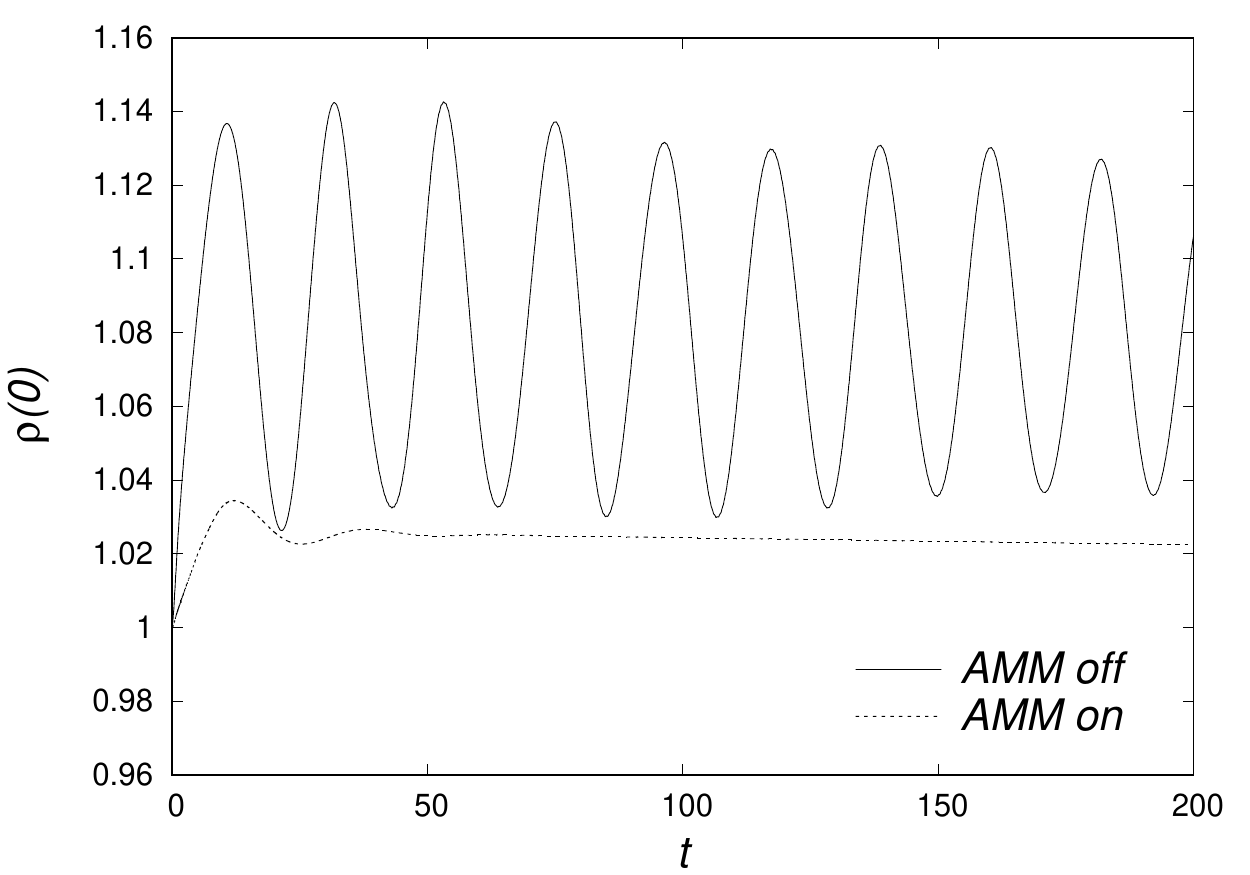}
\caption{Central density as a function of time for an equilibrium configuration using the xBEC code in unigrid and FMR modes, using the ADI evolution methods (top). Results were obtained using the AMM method in off and on modes (bottom).} 
\label{fig:Equilibrium}
\end{figure}

Physically, because the initial configuration is stationary in the continuum, it is expected that the density $|\Psi|^2$ remains time-independent during the evolution. Nevertheless, finite differences approximation and time integration introduce permanent truncation errors that perturb the wave function. The effect is that the configuration oscillates, in fact with specific mode frequencies (see e.g. \cite{Guzman2019}), whose amplitude should converge to zero when increasing resolution. The results using the xBEC code in unigrid and FMR modes appear in the first panel of Figure \ref{fig:Equilibrium}. This figure shows that the amplitude of density oscillation reduces by a factor of four when doubling resolution, indicating second-order convergence. On the other hand, the results for the AMM code in on/off modes appear in the second panel and shows also a similar convergent behavior. In the case of the AMM method, precise convergence is not expected since the resolution is not exactly double, but gradually increasing resolutions that depend on the domain coordinates 
{ Also notice that the oscillations are damped in the case of AMM-on mode, which is due to the spatially dependent resolution, that produces space-dependent discretization errors different in each part of the domain.}

\subsubsection{Problem B.2: evolution of a boosted equilibrium configuration} \label{subsect:B2}

The code needs to show the ability to simulate moving configurations. This is why a second test involves a boosted equilibrium configuration. The set up consists in redefining the wave function of an equilibrium configuration $\Psi_{eq}({\bf x},0)\rightarrow e^{-iv_z}\Psi_{eq}({\bf x},0)$, which produces the configuration to move along the $z$ direction with velocity $v_z$.

\begin{figure}
\includegraphics[width=7.5cm]{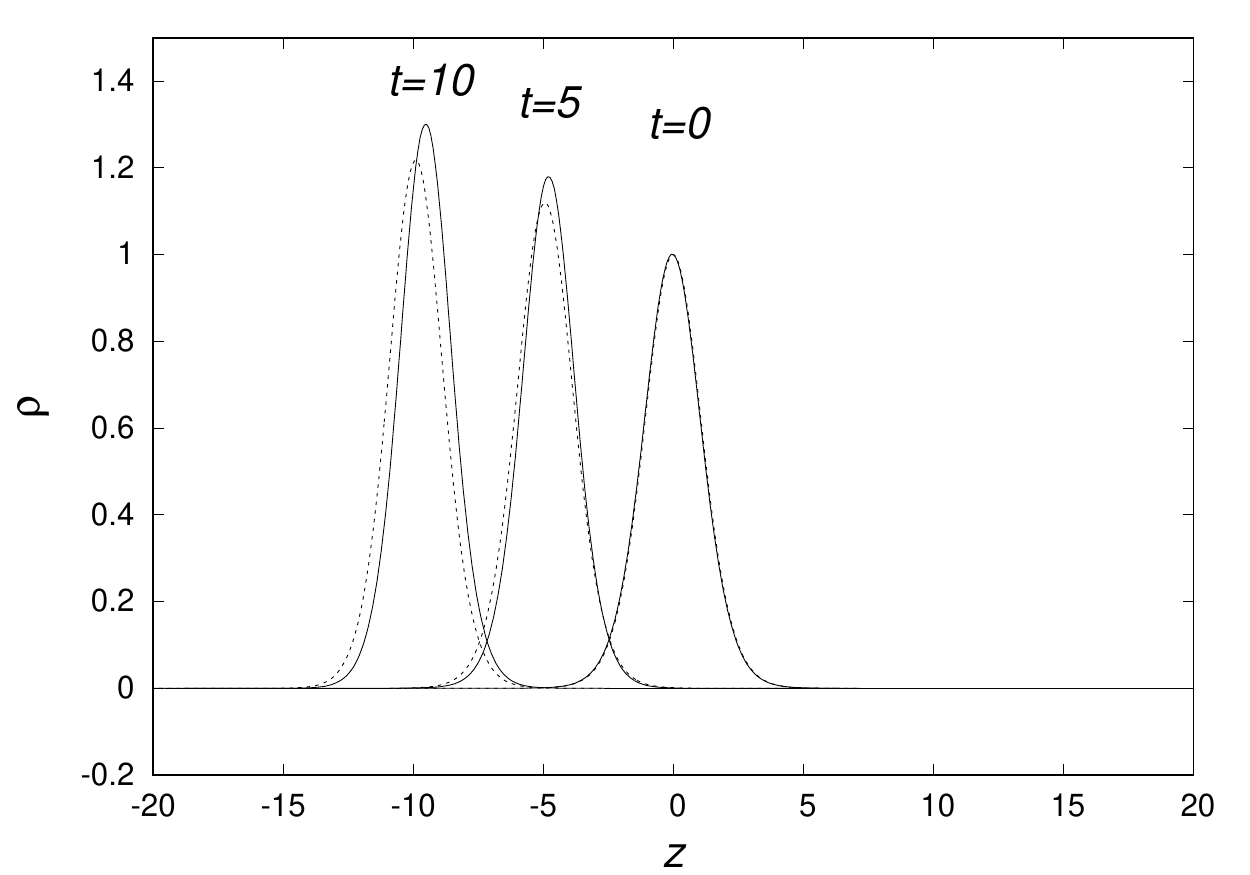}
\includegraphics[width=7.5cm]{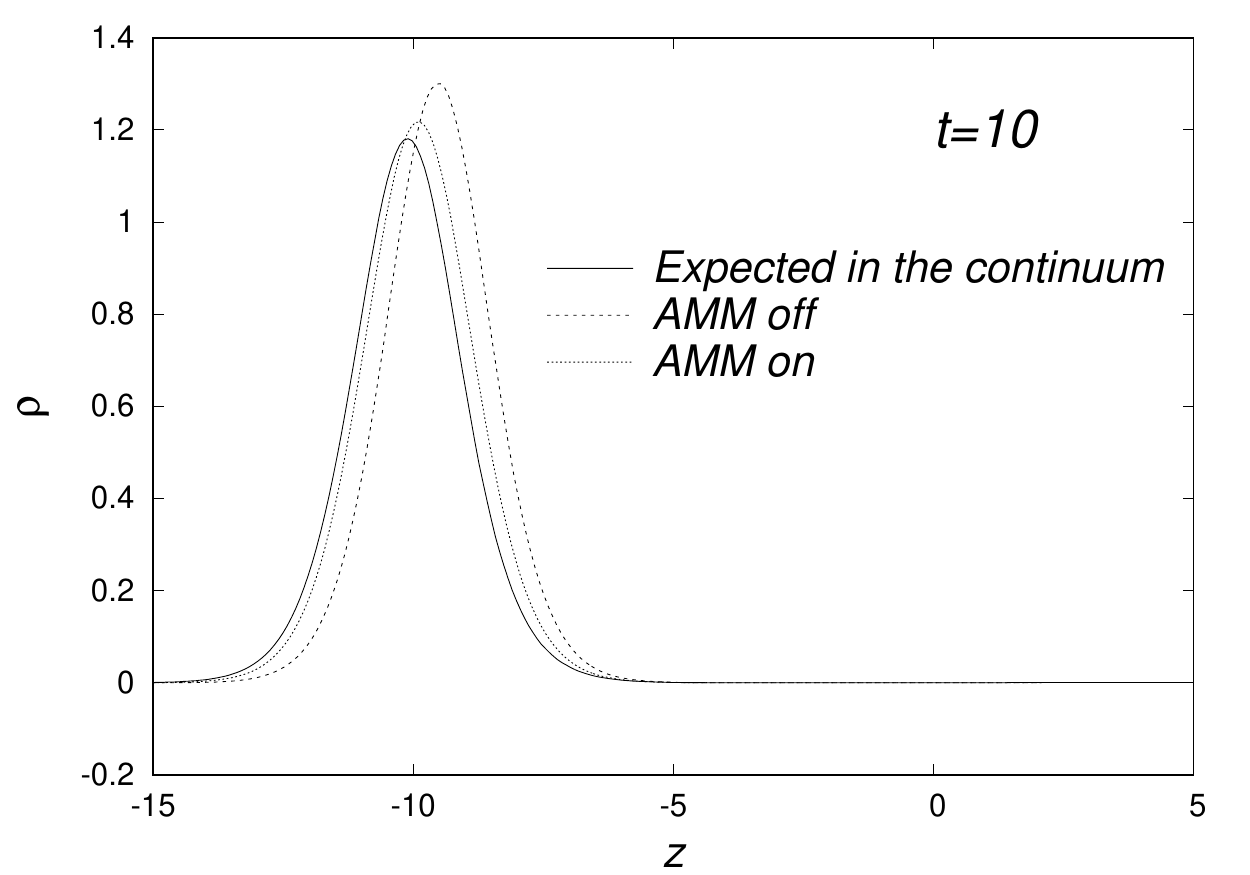}
\caption{{ (Top) Snapshots of density for the boosted configuration with $v_z=-1$ at time $t=0,5,10$ simulated with AMM in off and on modes. (Bottom) We show a zoom in of the solution at time $t=10$ and the estimated solution in the continuum limit calculated with a Richardson extrapolation with the xBEC code.}}
\label{fig:Boosted}
\end{figure}

For illustration, we use $v_z=-1$ and the configuration located initially at the coordinate origin. In Figure \ref{fig:Boosted} we show snapshots of the density $|\Psi|^2$ projected along the $z-$axis using the {AMM code in off and on modes. In order to know about the effects of refinement within the AMM code, we look closer at the snapshot at time $t=0$ and compare the results using the AMM-off and AMM-on modes with the expected solution in the continuum also in Figure \ref{fig:Boosted}. The AMM-off uses a constant resolution $\Delta xyz=0.4$ whereas the AMM-on uses a transformation that covers the range of resolutions from $\Delta xyz=0.4$ far from the blob to $\Delta xyz=0.2$ around the maximum density. The solution in the continuum is calculated as the Richardson extrapolation of solutions obtained with the xBEC in unigrid mode with resolutions $\Delta xyz=0.2$ and 0.1. The results show convergence toward the solution in the continuum when using the AMM-on mode.}. 

\begin{figure}
\includegraphics[width=2.75cm]{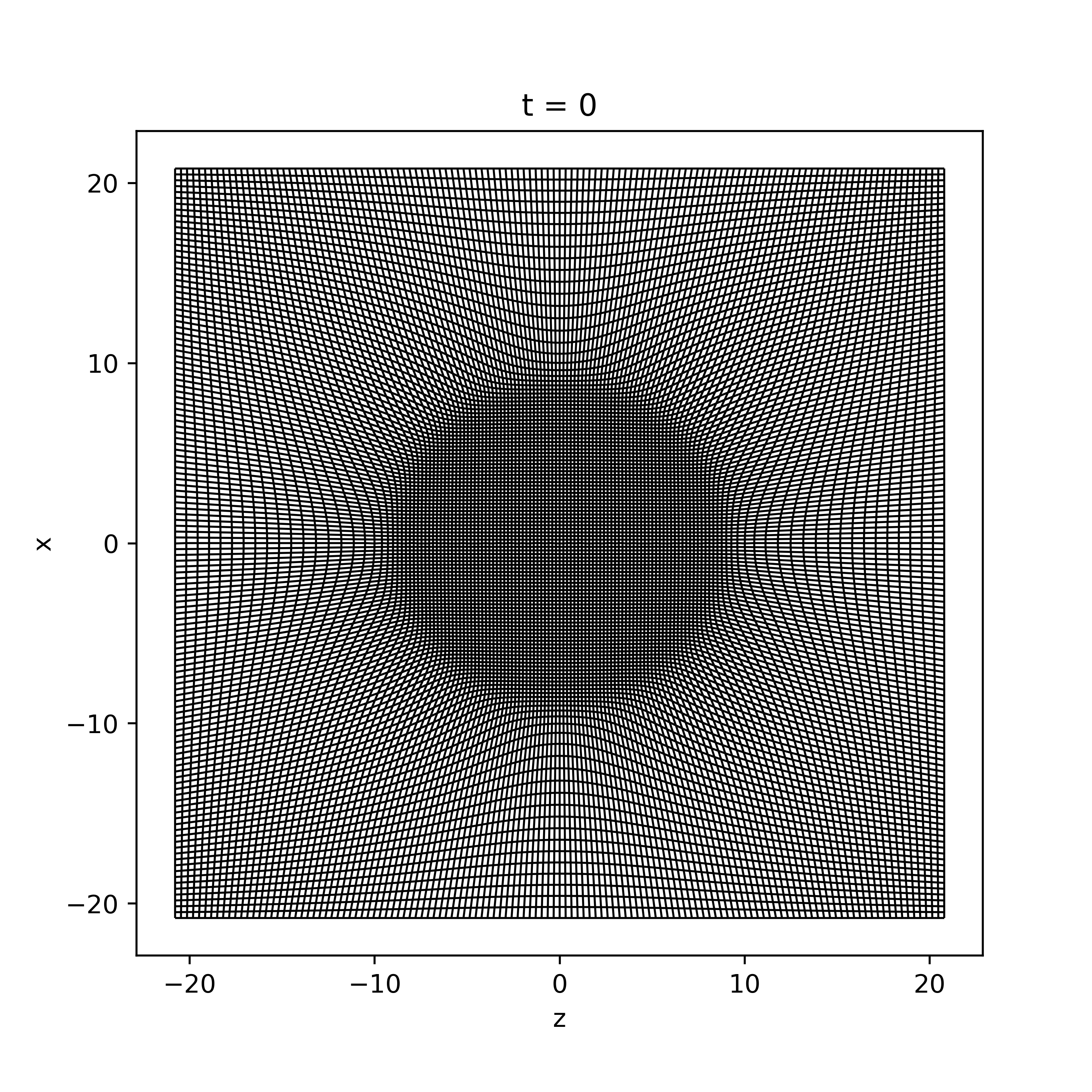}
\includegraphics[width=2.75cm]{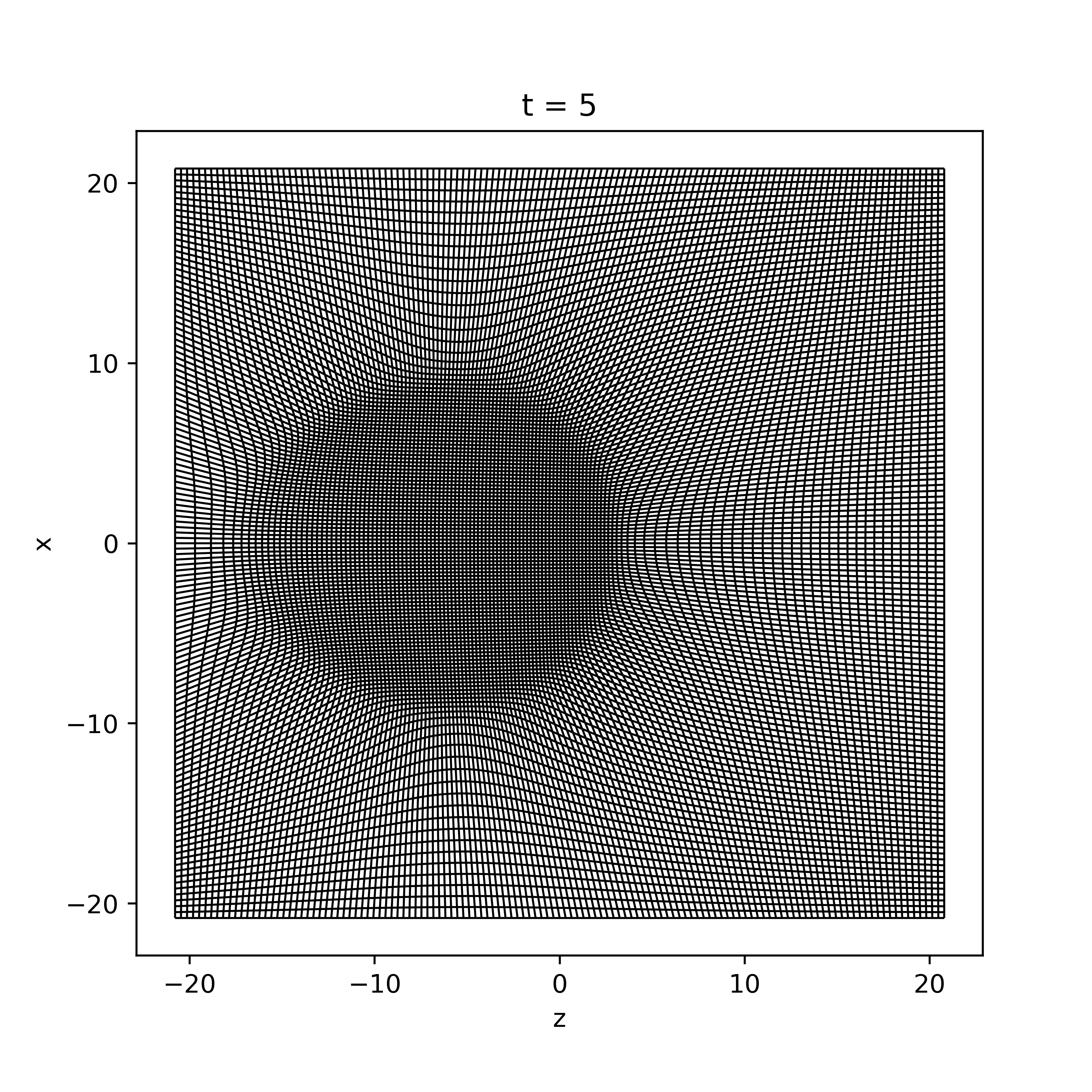}
\includegraphics[width=2.75cm]{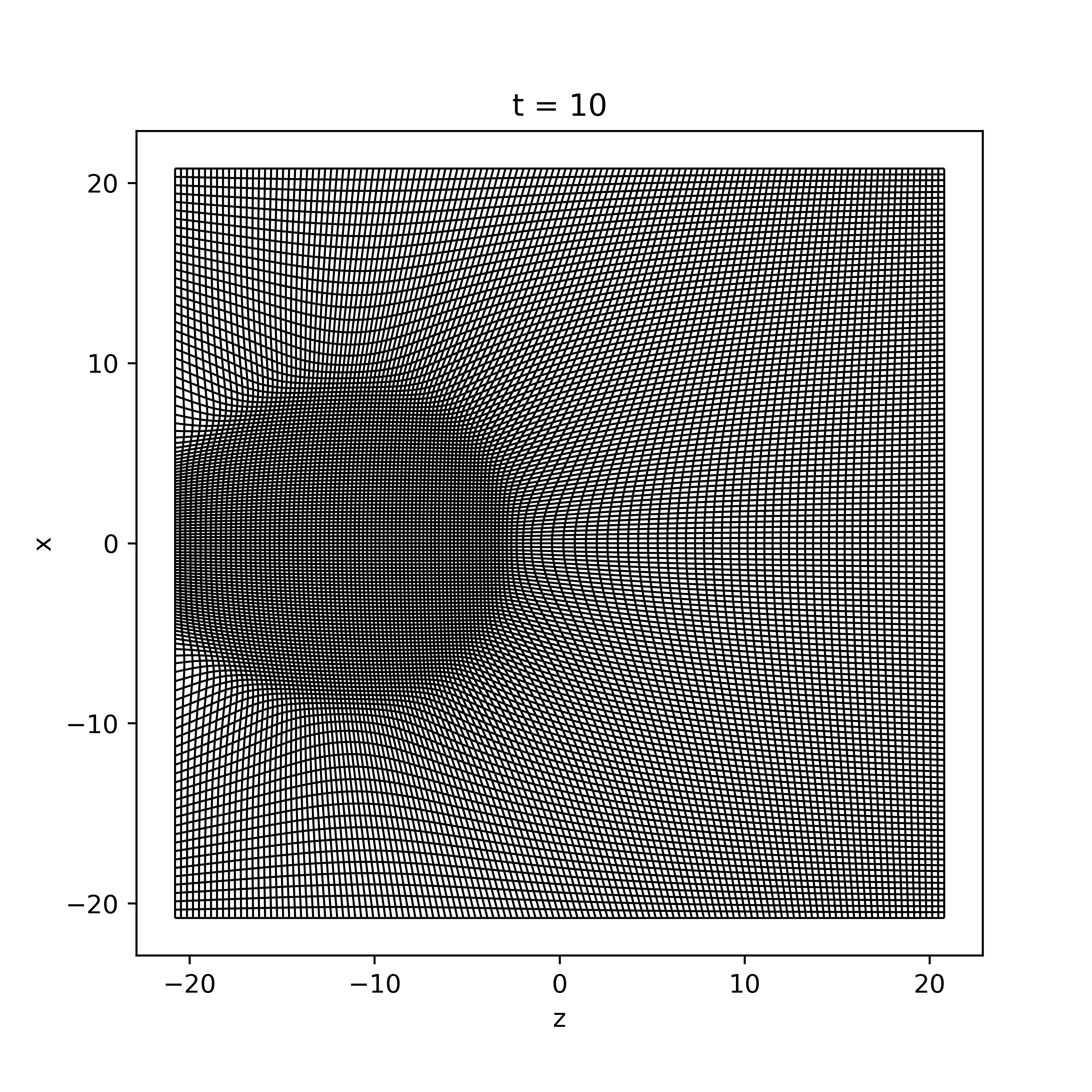}
\caption{Snapshots of the $zx-$plane at times $t=0,5,10$ of the physical numerical domain defined by the AMM method.} 
\label{fig:AMMBoosted}
\end{figure}

Most important for the AMM method is the evolution of the physical mesh illustrated In Figure \ref{fig:AMMBoosted}. Notice that the method designates higher resolution in regions with higher density gradients, which in this case is moving toward the left.

\subsubsection{Problem B.3: Evolution of a Binary Configuration}

\begin{figure*}
    \centering
    \includegraphics[width=15cm]{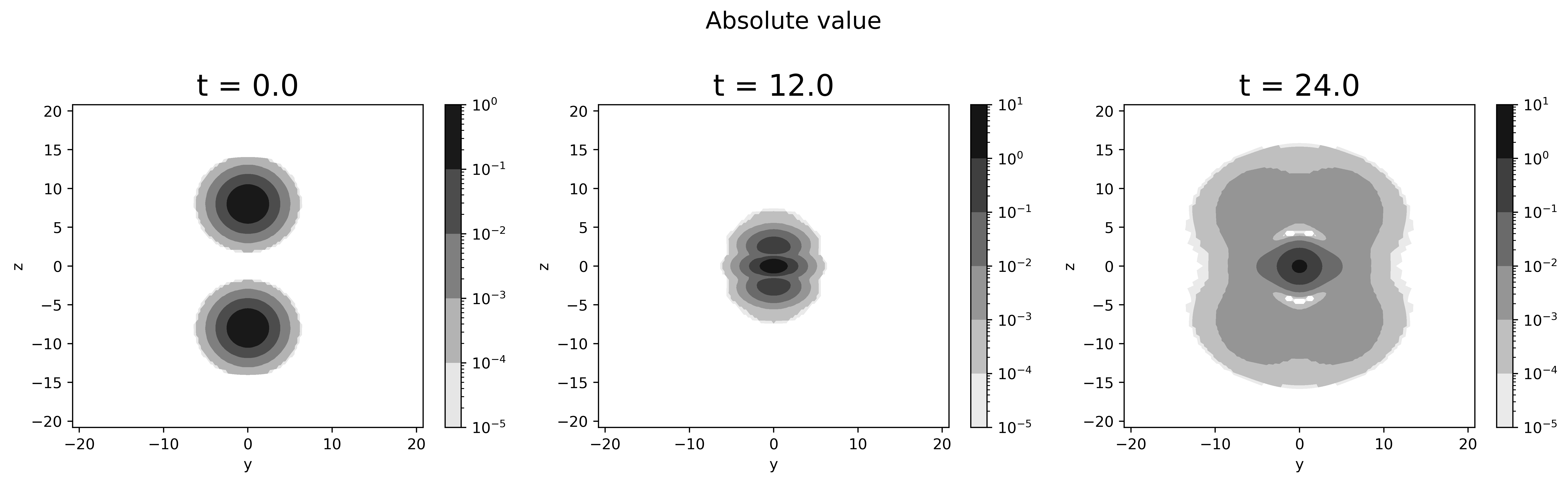}
    \caption{Projections of the density at three different times during the merger of the binary configurations on the $yz-$plane.} 
    \label{fig:Snaps2D}
\end{figure*}

\begin{figure}
\includegraphics[width=4.25cm]{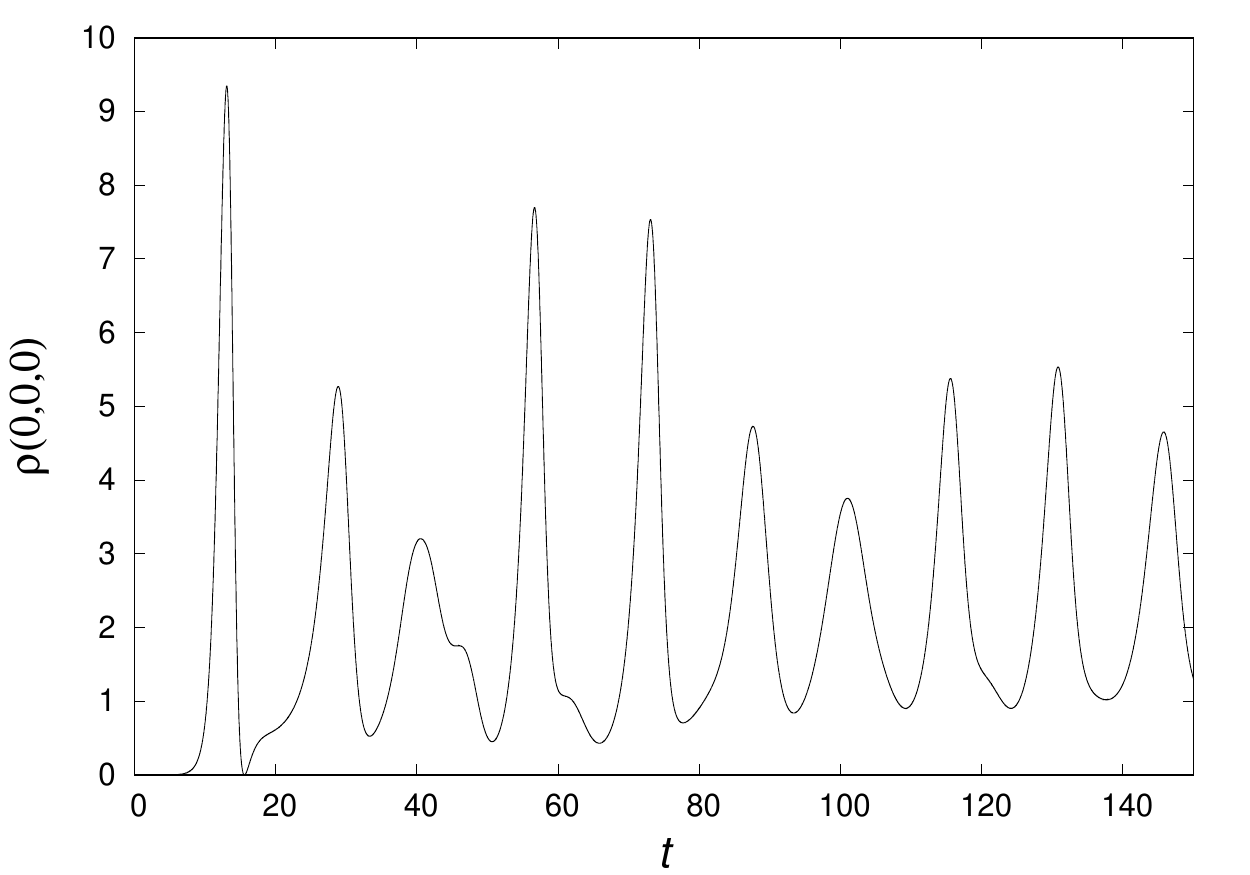}
\includegraphics[width=4.25cm]{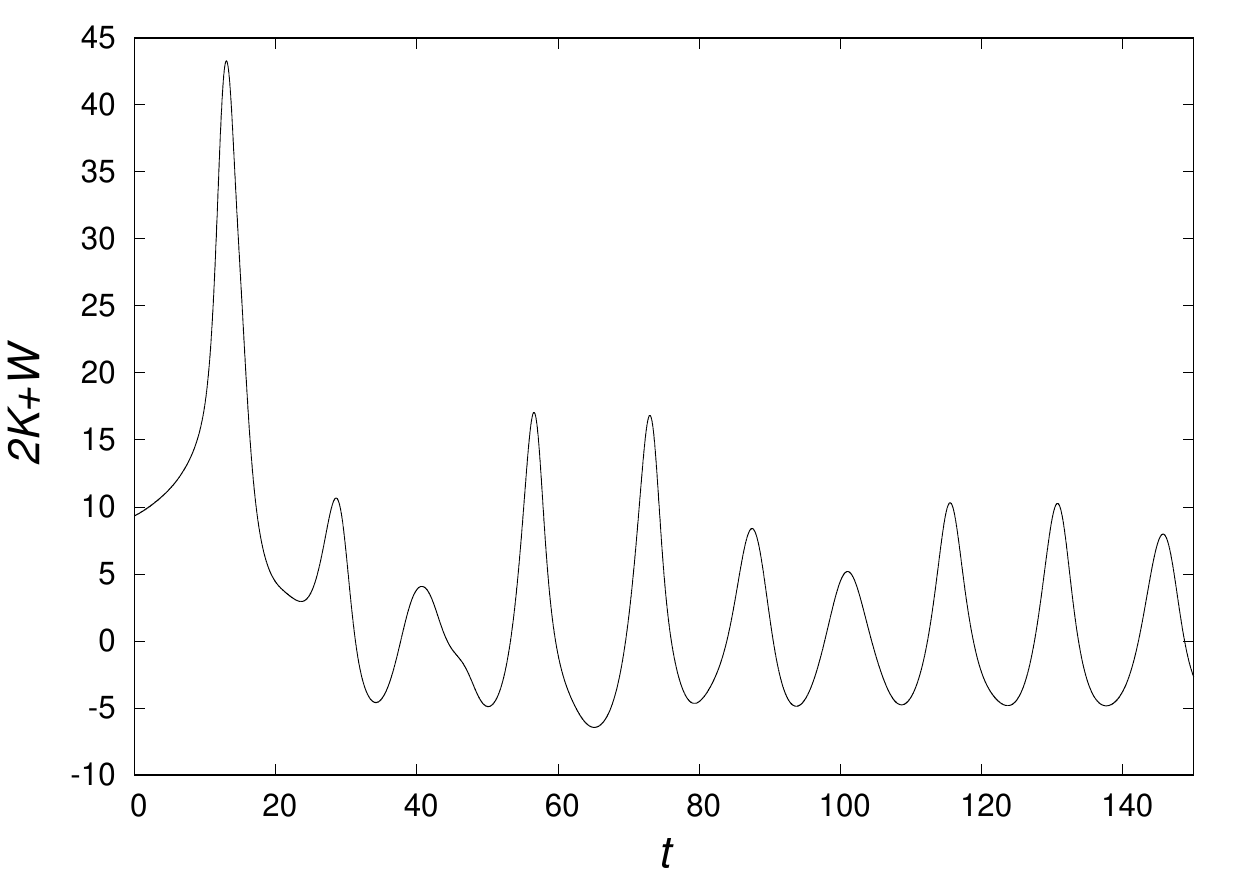}
\includegraphics[width=4.25cm]{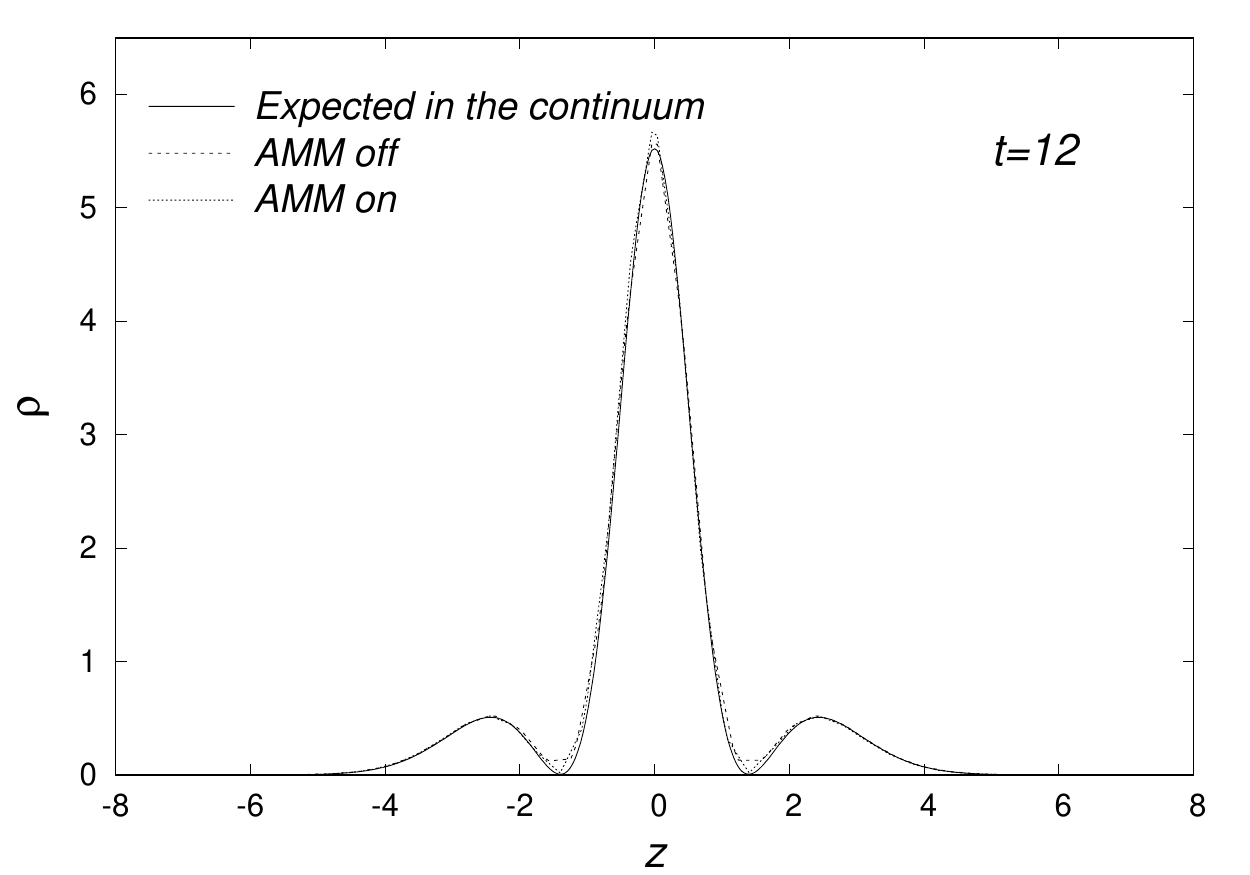}
\includegraphics[width=4.25cm]{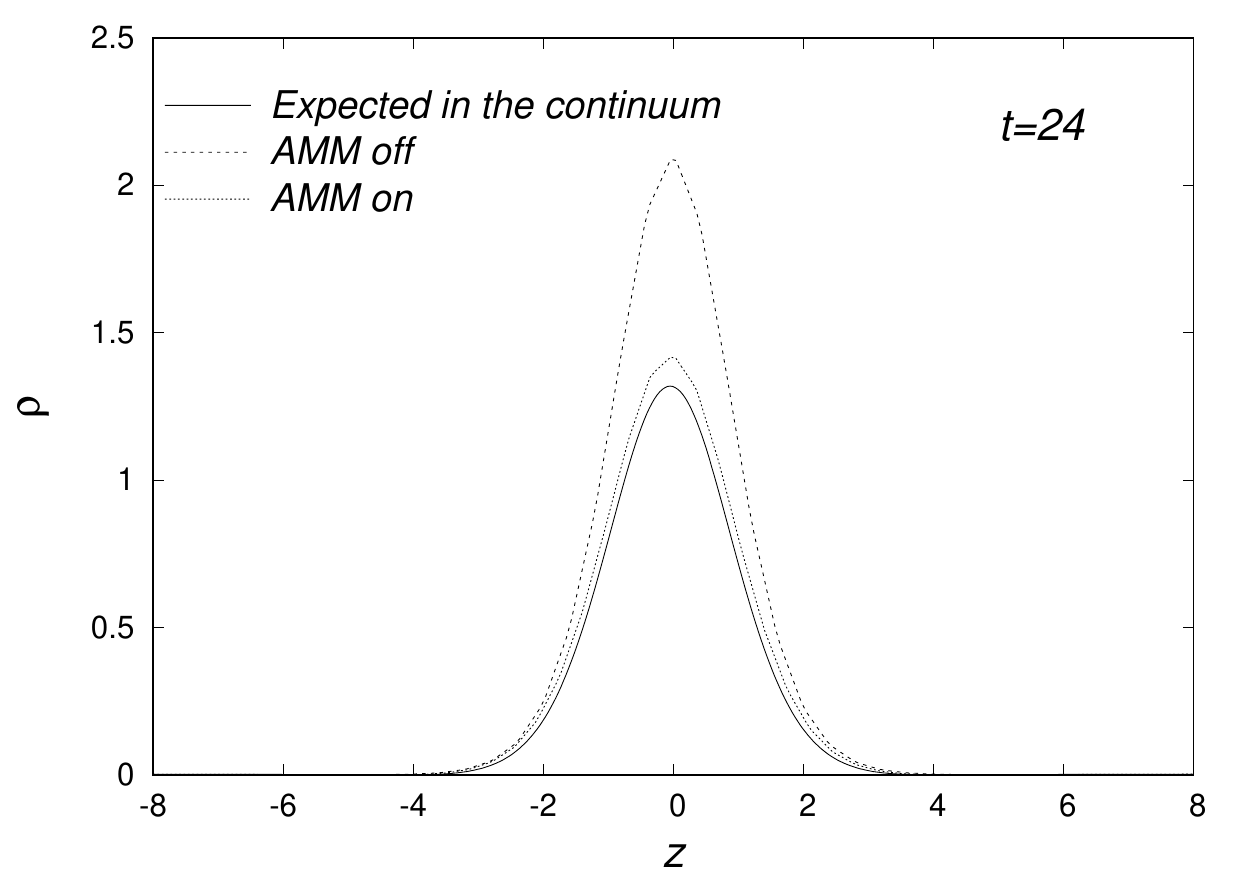}
\caption{The central value of the density and virialization function as a function of time for the head-on merger of two equilibrium configurations using the AMM method. {At the bottom we show the density using AMM-off and AMM-on converging to the expected solution at the continuum at two snapshots along the $z-$axis}.} 
\label{fig:AMMHeadOn}
\end{figure}

For the next step, we show the solution of the SP system for the frontal collision of two equal mass equilibrium configurations with head-on velocity $v_z =\pm 0.5$ launched from initial positions $z_0 = \pm 8$ along the $z-$axis. 
{ This case with $E<0$ corresponds to a merger where the two initial configurations fuse and for a final single blob. In Figure \ref{fig:Snaps2D} we show some snapshots of the projected density on the $yz-$plane. In Figure \ref{fig:AMMHeadOn} we show some diagnostics, including the time series of the central density of the system and the virialization function $2K+W$. Also shown is a convergence test of the density at two times. For this, we generated three runs with a different resolution with xBEC and produced the Richardson extrapolated solution to the continuum. The plots show that the solutions using AMM-off with uniform resolution $0.4$ and AMM-on with resolution from 0.4 to 0.2, the numerical solutions converge to the expected in the continuum.}

\subsection{Problem C: Embedded SP configurations inside a Soliton Halo}
\label{sec:ProbC}

The main purpose of this section is to show we can use our AMM code to simulate the evolution of a subhalo with the density profile of an equilibrium configuration, laying inside a background potential of a host halo whose gravitational potential dominates and eventually disrupts the subhalo . 

{Our main goal of considering this case is to explore a simple scenario reminiscent of SFDM halos laying in highly interacting astrophysical environments. For Problem C, we evolved an initially stationary configuration interacting gravitationally with a larger configuration with a fixed core-like density profile. Simulations of this kind are promising for studies of highly interacting galactic systems such as groups and clusters of galaxies which are scarce or even lacking so far within alternative scenarios to CDM such as the SFDM model. The complexity of phenomena involved in those systems is so high, that sophisticated numerical and physical methods are required in order to even accomplish a fair description of their dynamics and evolution. 
}.

With that goal in mind, we consider the host galaxy halo is spherical with density profile given as a soliton with a density profile prescribed from cosmological simulations as follows \cite{Yavetz:2021}

\begin{equation}
\rho_c(r)=\frac{0.019(m/10^{-22}\mathrm{eV})^{-2}(r_c/\mathrm{kpc})^{-4}}{\left[1+0.091\left(r/r_c\right)^2\right]^8}\label{eq:rhoc}
\end{equation}

\noindent where 

\begin{equation}
r_c=1.6\left(\frac{m}{10^{-22}\mathrm{eV}}\right)^{-1}\left(\frac{M_{vir}}{10^{9}\mathrm{M}_\odot}\right)^{-1/3}\mathrm{kpc},
\end{equation}

\noindent corresponding to a core size of approximately $2\,\mathrm{kpc}$. This density distribution sources a gravitational potential $V_{bg}$ that is kept fixed during the evolution.

The host halo produces a fixed gravitational potential that is plugged into the Schr\"odinger equation. This potential arises from solving the Poisson equation $\rho_c(r)$ in (\ref{eq:rhoc}) at the initial time and plays the role of a gravitational trap that confines the substructure into a bound region. 

\begin{figure}
\includegraphics[width=8cm]{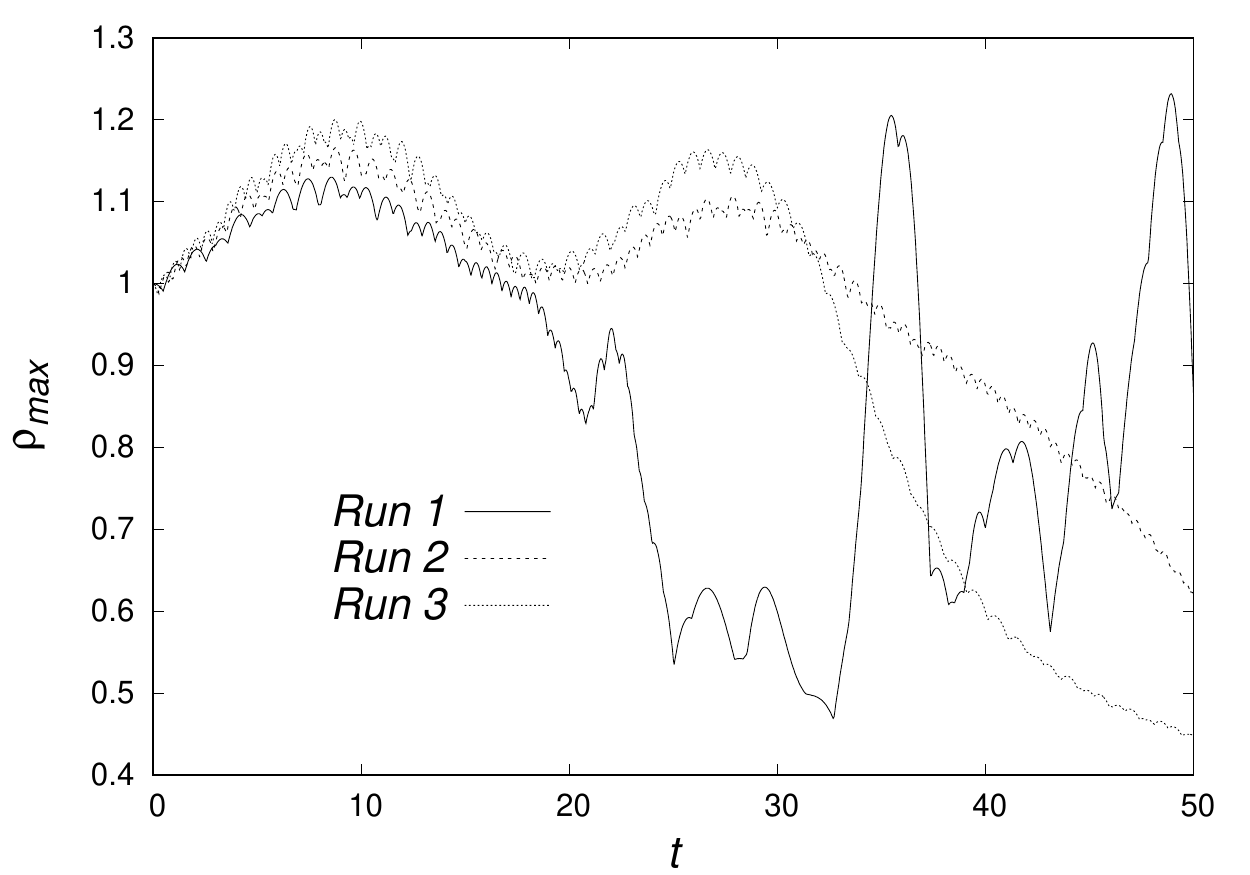}
\includegraphics[width=8cm]{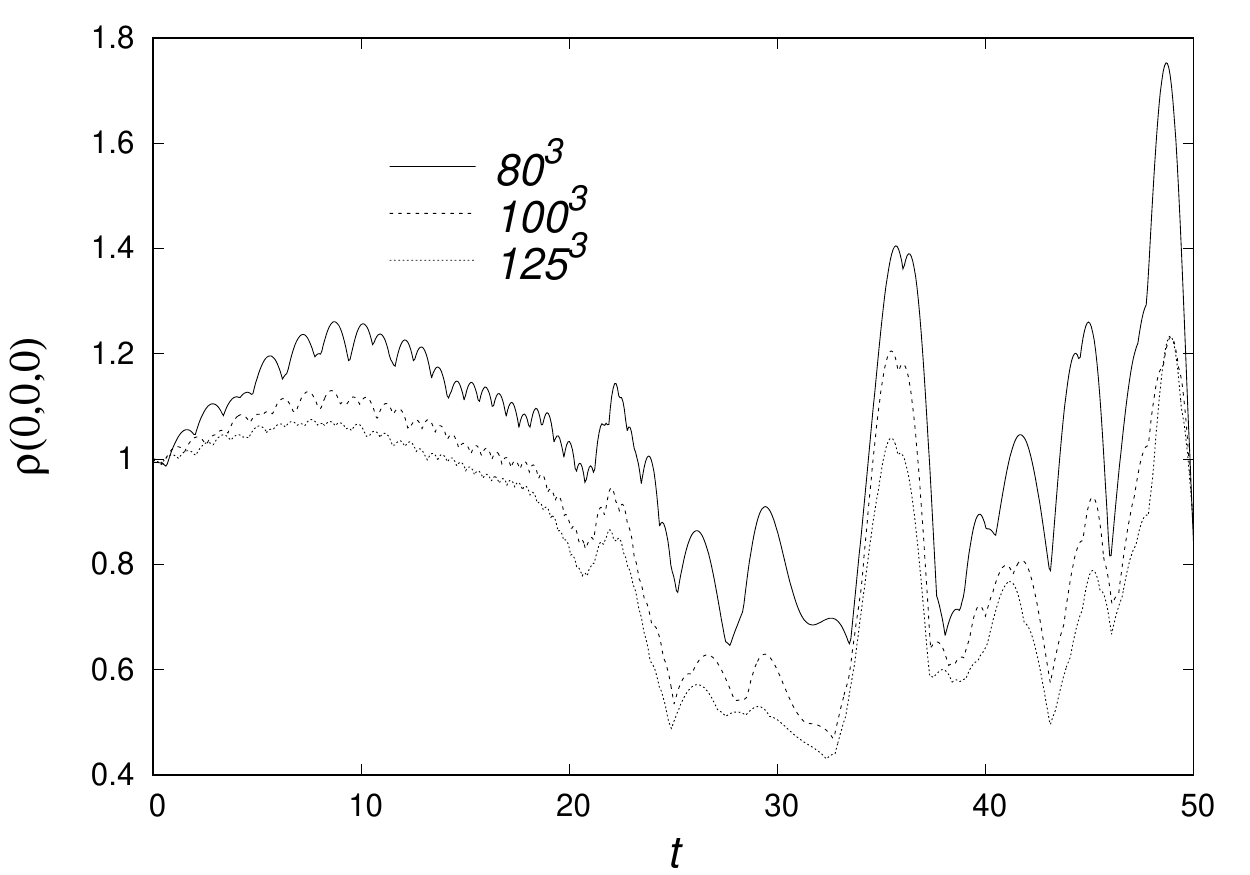}
\caption{{ (Top) Maximum of the subhalo density $|\Psi|^2$ as a function of time, for different initial conditions. (Bottom) Self-convergent behavior of the maximum density for Run 1 using three resolutions with resolution factor 1.25=125/100=100/80.}}
\label{fig:LargePlusSubstruct}
\end{figure}

\begin{figure*}
\includegraphics[width=15cm]{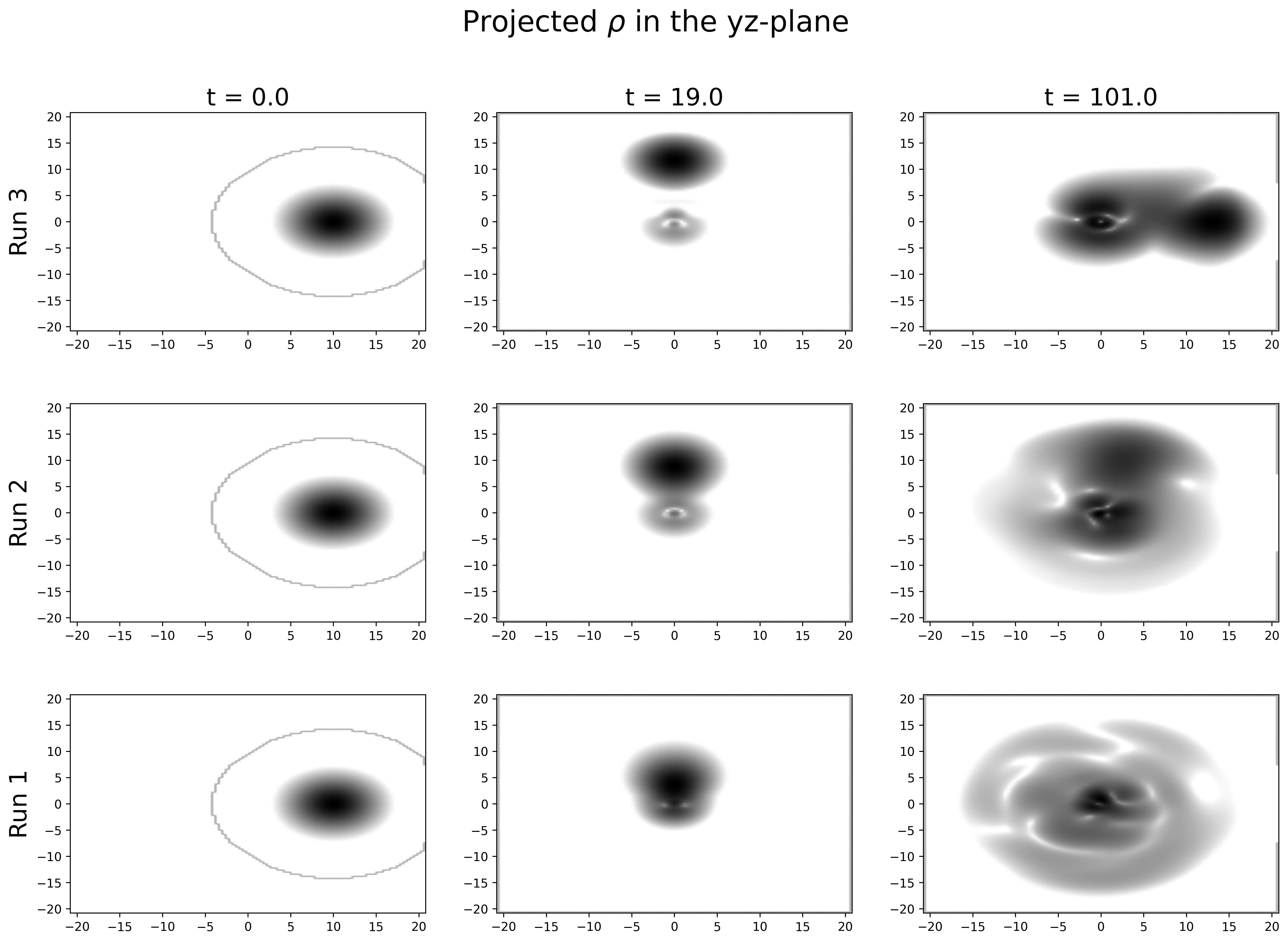}
\caption{Snapshots of the subhalo density projected on the $yz-$plane at times $t=0,19,101$ from left to right in different simulations. For reference, the center of the host galaxy lies at the coordinate origin.}
\label{fig:Disrupt}
\end{figure*}

In our case of study, we consider a mass of the host halo that is larger than the mass of the subhalo by a factor of two. Since it is of our interest to study the disruption effect suffered by the subhalo due to the interaction with the host, we set the initial conditions such that the center of the host halo stays fixed at the origin of coordinates, whereas the subhalo configuration is initially centered at $\vec{x}(0)=(0,10,0)$ in code units (corresponding to $\vec{x}(0)=(0,19.2,0)\,\mathrm{kpc}$, using $v\sim 10 km/s$ as the average velocity of bosons in the condensate \cite{Hui:2016} in order to calculate the de Broglie length), with initial velocity $\vec{v}(0)=(0,0,-v_0)$. The initial velocity magnitude $v_0$ is the only parameter that is varied in the three different simulations. In order to set the values of $v_0$ we take as reference the value $v_{ref}=\sqrt{\frac{M_c(r)}{r}}$ and $M_c(r) = \int_0^r \rho_c(r')r'^2 dr'$. The idea is that with this initial velocity the trajectory of the center of the subhalo corresponds to a nearly circular orbit on the $yz-$plane.

The three runs correspond to different values of the initial velocity of the subhalo defined as follows: 

\begin{itemize}
    \item[Run 1] $v_0 = v_{ref}-0.2\simeq 0.4$. In this case it is expected that the gravitational force dominates over the centripetal force. 
    \item[Run 2] $v_0 = v_{ref}\simeq 0.6$. The subhalo would move in a nearly circular trajectory. Due to tidal effects, the spherical shape would be destroyed and part of the mass of the subhalo would be torn apart whereas the other part would be accreted towards the center of the potential. 
    \item[Run 3] $v_0 = v_{ref}+0.2\simeq 0.8$. The subhalo would tend to escape from the host potential and eventually its mass will be lost through the boundaries. 
\end{itemize}
  
The settings for the AMM-on mode of the code are the same as in Problem B (see \ref{sec:tests_SP}). The domain for all our simulations is given by a cubic box with edges placed at: $[-20.8,20.8]^3$ which is discretized into a mesh containing $N=104^3$ points.

Figure \ref{fig:LargePlusSubstruct} shows the maximum density of the subhalo , { for Runs 1, 2 and 3 as a function of time. Also in Figure \ref{fig:LargePlusSubstruct} we show the self convergent behavior for Run1 of the maximum of density, using three runs with successive resolutions with resolution factors of 1.25.} 

{We considered the time series of the maximum density of the subhalo as an indicator of the extent of deformation of the subhalo. The behavior of density maximum indicates that the subhalo } tends to remain closer to the initial value as the angular momentum is bigger and also oscillates with larger amplitude. In conclusion, the larger the angular momentum is, the more deformed results of the subhalo. Finally, in Figure \ref{fig:Disrupt} we illustrate the evolution of the subhalo density for the three runs, projected on the $yz-$plane. 
{ Interestingly, as clearly can be seen in snapshots shown in Figure \ref{fig:Disrupt}, we can distinguish two possible fates of the subhalo, for run 1 since the angular momentum is sufficiently small, a significantly large fraction of the mass remains as a bounded configuration infalling towards the host center, while the rest of the initial mass is ripped away from the subhalo and either scatters around the host potential well or gets away from it. In conclusion, in the last scenario, the disruption effect on the subhalo results in a deformation and a reduction in mass. In contrast, in the other side limit, simulated in run 3, the subhalo holds large enough angular momentum to dissolve the subhalo and part of its mass is kidnapped by the host and the other fraction runs away from it. }

\section{Final comments}
\label{sec:comments}

We have presented the use of the AMM method applied to Initial Value Problems associated with the Schr\"odinger equation, under various scenarios. Being our main interest the application to the dynamics of ultralight bosonic dark matter at local scales, we study cases with localized distributions of bosons where gravity plays a role. 

{We consider 2 simple cases which have been extensively studied in the literature. Problems A and B served as tests to our AMM code. Specifically, we tested the numerical performance and precision of our code (both AMM-on and off options) by comparing its outputs to well-known solutions. However, we did not explore any new physics since this has been addressed before. }

For these problems, we have shown the ability of the AMM method to provide adaptability of the numerical resolution to desired high-density regions in the physical domain, where simulations are expected to be more accurate.

We have shown that the method can handle scenarios involving the dynamics of dark matter configurations, and expect the implementation of this method contributes to the analysis of at least galaxy scale phenomena related to ultralight bosonic dark matter.

{ Particularly, we studied Problem C as a simple but fairly complex scenario which brings up some insights for studying highly interacting galactic systems such as groups and clusters of galaxies that are scarce or even lacking so far within alternative scenarios such as the SFDM model. The complexity of phenomena involved in those systems is so high, that sophisticated numerical and physical methods are required in order to even accomplish a fair description of their dynamics and evolution. 
Our main achievement regarding problem C was to track the evolution of a configuration playing the role of a galactic-sized SFDM-subhalo suffering a disrupting effect produced by the gravity of a host halo.
Out main conclusions regarding this problem are: (1) The largest the angular momentum the stronger disruption of the subhalo and (2) there are two possible fates of the subhalo after being disrupted: it remains bounded or it is dissolved into the host. 
 }


\section*{Acknowledgments}
E.M.V and J.L.S acknowledge CONACYT for PhD fellowships. 
A.A.L acknowledges the financial support from VIEP-BUAP.
FSG acknowledges support from grants CIC-UMSNH-4.9, CONACyT Ciencias de Frontera Grant No. Sinergias/304001. 
Runs were carried out in the Big Mamma cluster at the Laboratorio de Inteligencia Artificial y Superc\'omputo, IFM-UMSNH and at the Laboratorio Nacional de Supercómputo del Sureste de México under the institutional project: "Estudio de Modelos de Materia Oscura a partir de datos Cosmológicos Diversos".


\appendix

\renewcommand\thefigure{\thesection.\arabic{figure}}    
\section{Accuracy and computational cost}
\setcounter{figure}{0}

We present a measure of the trade-off between accuracy and computational cost for the AMM-on and AMM-off modes of our code.
The Schr\"odinger-Poisson system is a constrained evolution system composed by an elliptical constraint and a parabolic-type evolution equation. 
The evolution from time $t^n$ to time $t^{n+1}$ for a variable $\Psi$ is given by an evolution operator $\Psi^{n+1}=A\Psi^{n}$ for all points in the spatial domain, where $A = \left(\mathbb{I}+\mathbb{M}\right)$ uses an explicit discretization of the evolution equation.


{ On the other hand, Poisson equation \eqref{eq:poisson} is an elliptic equation, whose solution requires an iterative algorithm, in this work we used the multigrid method, in which  Poisson equation is written as a linear system of equations (LSE) for the potential $Au_m=b$, where $u_m = V^{n}_m$ is the m-th step in the recursion for the n-th iterative time step, $b = |\psi^{n}_{i,j,k}|^2$ is the source and $A$ corresponds to the matrix representation of the Laplacian operator. The above is applied recursively until the infinity norm of the error $Err = |b - A u_i|_{\infty}$ of the residual crosses a threshold close to zero. }

Due to the fact that the solution of Poisson equation requires the solution of an LSE problem  several times in each time step, it is the most expensive operation in computational resources terms. For that reason we consider Poisson equation to analyze computational cost, in the a) AMM ON and b) AMM OFF scenarios. 

\begin{figure}
    \centering
    \includegraphics[width = 8 cm]{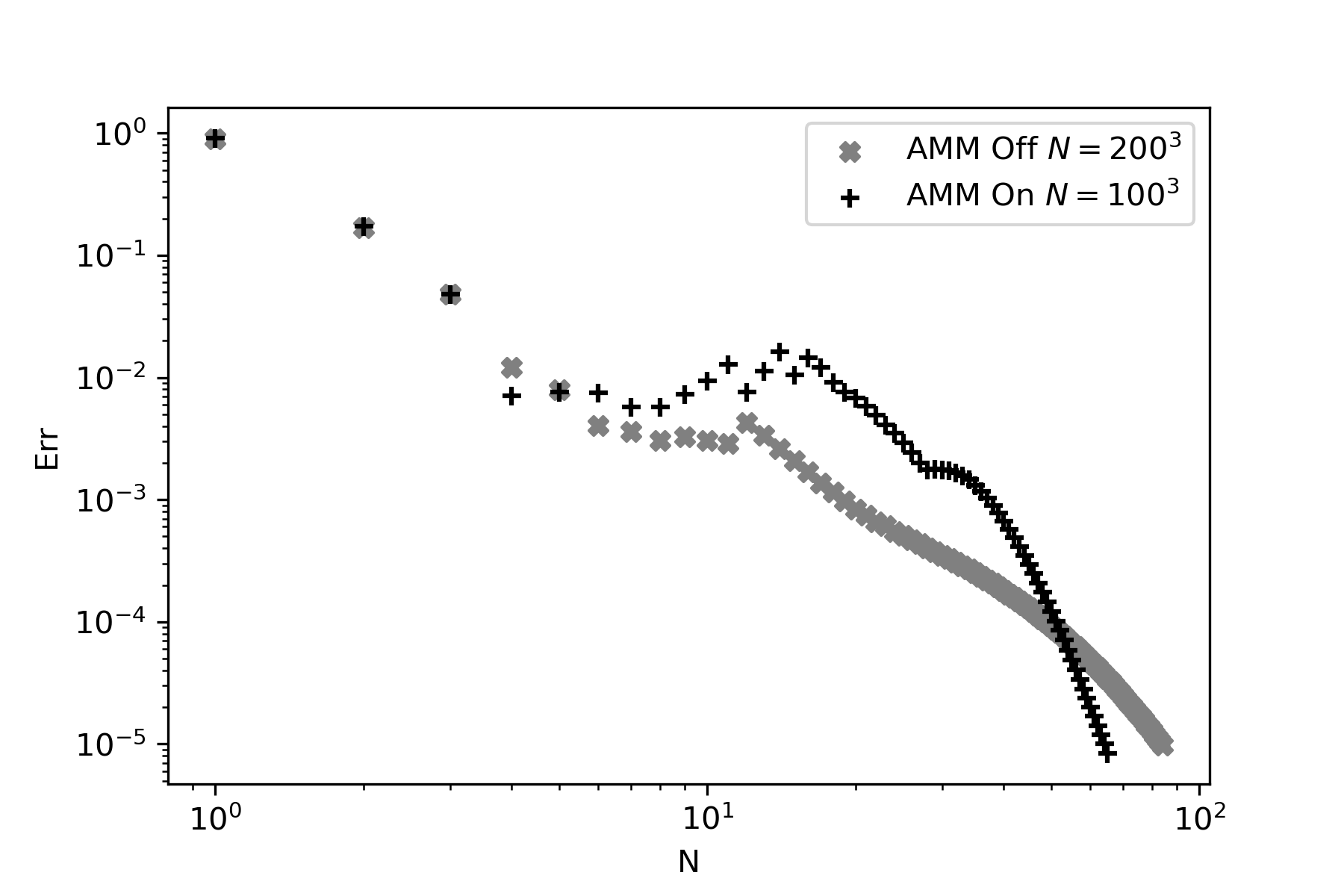}
    \caption{{The error of the solution of Poisson equation vs number of V-cycle iterations. The convergence of the solution of Poisson's equation is shown. For the threshold $10^{-5}$ the use of only $100^3$ cells with AMM-on shows to be more efficient that AMM-off.}}
    \label{fig: PoissErr}
\end{figure}

We calculate the computational cost using the convergence rate metric for the multigrid method taking a truncation error $Err = 10^{-5}$. A stationary configuration was studied within a box $[- 20,20 ]^3$ in both scenarios, the first one with $N=100^3$ for \textbf{a)} achieving a central resolution of $\Delta_x \sim 0.2$, and the second one with $N= 200^3$ points for \textbf{b)} also with a resolution of $\Delta_x = 0.2$. In both cases the initial guess is the field zero. 
The results are shown in figure \ref{fig: PoissErr}, where we observe that scenarios (a) and (b) the number of iterations needed for convergence are of the same order. However, for the latter case, the number of operations performed by the CPU is $2^3 = 8$ times bigger than in the AMM-ON case. This allows one to  quantify the difference of using a logical domain discretized with $100^3$ cells instead of $200^3$ cells with similar accuracy in the refined region.

\bibliography{bibliografia}

\end{document}